\documentclass[journal,compsoc]{IEEEtran}
\usepackage{iftex}

\usepackage{amsmath,amsfonts}

\usepackage{graphicx}
\usepackage{stfloats}

\usepackage{tabularx}
\usepackage{array}
\usepackage{multirow}
\usepackage{multicol}

\usepackage{algorithm}
\usepackage{algorithmic}

\usepackage{enumitem}
\usepackage{textcomp}
\usepackage{verbatim}

\usepackage{xcolor}
\usepackage{pifont}
\usepackage{bbding}

\usepackage{cite}
\usepackage{url}
\usepackage[strings]{underscore}

\usepackage[caption=false,font=normalsize,labelfont=sf,textfont=sf]{subfig}

\usepackage{soul}
\usepackage{pifont}
\soulregister\cite7
\soulregister\ref7

\ifXeTeX
\usepackage{fontspec}
\defaultfontfeatures{Ligatures=TeX}
\newcommand{\TLGyrePath}{/usr/local/texlive/2025/texmf-dist/fonts/opentype/public/tex-gyre/}
\IfFileExists{/usr/local/texlive/2025/texmf-dist/fonts/opentype/public/tex-gyre/texgyretermes-regular.otf}{}{%
\renewcommand{\TLGyrePath}{c:/texlive/2025/texmf-dist/fonts/opentype/public/tex-gyre/}%
}
\fi

\usepackage{easyReview}

\newcommand{\cnum}[1]{\ding{\numexpr171+#1\relax}}

\newcolumntype{Y}{>{\centering\arraybackslash}X}

\begin{document}

\title{HALF: Hollowing Analysis Framework for Binary Programs with Kernel Module Assistance}

% \author{
%     \IEEEauthorblockN{Letian Sha$^{1}$, Jiaye Pan$^{1}$}
%     \IEEEauthorblockA{$^{1}$School of Computer, University of Posts and Telecommunications, Nanjing 210003, China.}
% }

\author{Zhangbo Long, Letian Sha$^{\ast}$, Jiaye Pan$^{\ast}$, Haiping Huang, Dongpeng Xu, Yifei Huang, Fu Xiao%
\thanks{Letian Sha and Jiaye Pan are co-corresponding authors.}
\thanks{Zhangbo Long, Letian Sha, Jiaye Pan, Haiping Huang, Yifei Huang, and Fu Xiao are with the School of Computer Science, Nanjing University of Posts and Telecommunications, Nanjing, Jiangsu, 210023, China.
E-mail: 2023040507@njupt.edu.cn, ltsha@njupt.edu.cn, panjy@njupt.edu.cn, hhp@njupt.edu.cn, yifei_huang@bupt.cn, xiaof@njupt.edu.cn.}
\thanks{Dongpeng Xu is with The University of New Hampshire, Durham, New Hampshire 03824, United States.
E-mail: dongpeng.xu@unh.edu}

}

% \author{Letian Sha, Jiaye Pan}
% \address{School of Computer, University of Posts and Telecommunications, Nanjing 210003, China.}
%\thanks{Manuscript received April 19, 2021, revised August 16, 2021.}

% \markboth{Journal of \LaTeX\ Class Files,~Vol.~14, No.~8, August 2021}%

% \IEEEpubid{0000–0000/00$00.00~\copyright~2021 IEEE}
% Remember, if you use this, you must call \IEEEpubidadjcol in the second

\IEEEtitleabstractindextext{
\begin{abstract}
Binary program analysis represents a fundamental pillar of modern system security. Fine-grained methodologies like dynamic taint analysis still suffer from deployment complexity and performance overhead despite significant progress. Traditional in-process analysis tools trigger severe \textbf{address-space conflicts} that inevitably disrupt the native memory layout of the target. These conflicts frequently cause layout-sensitive exploits and evasive malware to deviate from their intended execution paths or fail entirely. This paper introduces \textbf{HALF} as a novel framework that resolves this fundamental tension while ensuring both analysis fidelity and practical performance. HALF achieves high-fidelity address-space transparency by leveraging a kernel-assisted process hollowing mechanism. This design effectively eliminates the observation artifacts that characterize traditional instrumentation tools. We further mitigate the synchronization latency of decoupled execution by implementing an exception-driven strategy via a lightweight kernel monitor. Extensive evaluation of a Windows-based prototype demonstrates that HALF maintains superior performance compared to conventional in-process baselines. HALF also provides unique capabilities for deconstructing complex, stealthy threats where existing frameworks fail to maintain execution integrity.
\end{abstract}

}
\maketitle

\section{Introduction}
Binary program analysis plays a critical role in modern cybersecurity practice.
It supports vulnerability triage and root-cause investigation for Common Vulnerabilities and Exposures (CVE) reports, enables evidence-chain reconstruction in advanced persistent threat incident response, and assists vendors and defenders in validating the security impact and correctness of software patches~\cite{martinliras_2021_feature,smallissery_2023_demystify,lstone_2023_no}.
As attackers continuously evolve stealth techniques and defenders must respond under time pressure, analysts increasingly rely on analysis methods that operate directly on binaries to obtain a reliable, deployment-close understanding of program behavior~\cite{schakkaravarthy_2019_a}.
These realities place binary-centric reasoning at the center of defensive workflows and make binary program analysis a foundational capability for both security operations and research.

To meet real-world demands, binary analysis must move beyond coarse-grained behavioral observations and provide fine-grained semantic evidence.
In adversarial settings, obfuscation and behavior-hiding strategies make high-level hooking insufficient for understanding intent and key data dependencies~\cite{bcheng_2023_on,ryang_2020_ratscope}.
On Windows, malicious or suspicious programs may invoke sensitive operations via component object model and remote procedure call, thereby weakening coverage of approaches that focus on direct Application Programming Interface (API) interception~\cite{fbarr-smith_2021_survivalism,rhund_2016_the}.
Fine-grained dynamic techniques, including dynamic Data Flow Tracking (DFT) and Dynamic Taint Analysis (DTA), have therefore become essential because they can expose data provenance and propagation at instruction-level granularity~\cite{dzhu_2010_tainteraser,pkemerlis_2012_libdft,jschwartz_2010_all,kjee_2012_a}.
However, achieving such fidelity requires pervasive instrumentation, which introduces nontrivial performance overhead and deployment friction, and motivates us to revisit the instrumentation-induced challenges in DTA.

\begin{figure}[!htbp]
\centering
\includegraphics[scale=0.4]{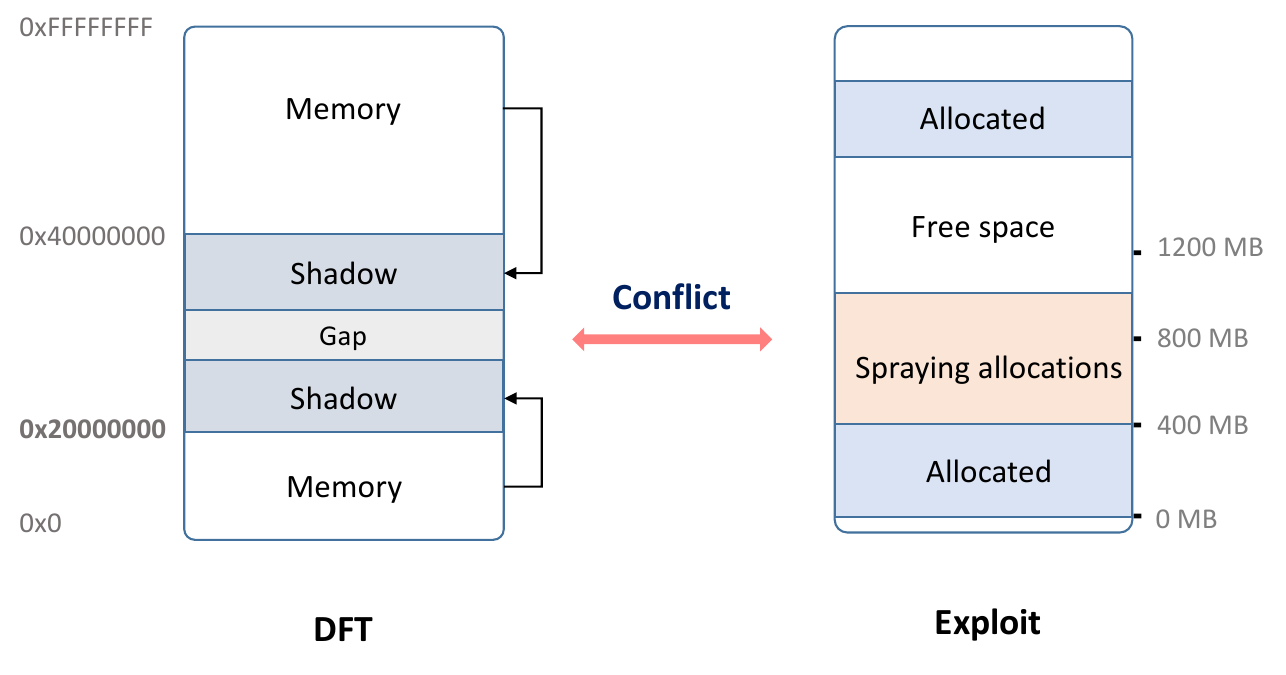}
\caption{Example of process space memory allocation conflict.
This figure illustrates memory usage conflicts between typical DFT tools and specific exploit programs.
Left side (DFT): shows that analysis tools (such as libdft) pre-allocate large memory regions (gray area near 0x20000000) in process space to maintain shadow memory for taint analysis.
Right side (Exploit): shows that certain exploit programs attempt to use heap spraying techniques to arrange memory in specific fixed address ranges (approximately 400MB - 800MB).
Conflict: pre-allocation by analysis tools occupies memory addresses required by exploit programs, causing conflicts that prevent exploit code from executing as expected or lead to abnormal program behavior.}
\label{fig:psmal}
\end{figure}

\textbf{Challenge.} Fine-grained dynamic binary analysis typically depends on runtime instrumentation and auxiliary analysis infrastructure, which jointly introduce nontrivial overhead and deployment cost.
Existing solutions commonly fall into two categories: virtualization-based approaches that incur substantial system-level overhead, and user-space in-process frameworks (e.g., DBI and DFT) that offer better flexibility but must embed analysis state into the target process.
In practice, these designs expose two recurring obstacles when analyzing real-world, layout-sensitive binaries: \textbf{(i) address-space interference that leads to allocation conflicts}, and \textbf{(ii) high synchronization/Inter-Process Communication (IPC) overhead in decoupled designs}.

First, address-space interference undermines analysis transparency.
Many DFT/taint frameworks reserve large contiguous regions in the target process to maintain shadow memory.
As illustrated in Fig.~\ref{fig:psmal}, libdft-style designs pre-allocate on the order of hundreds of megabytes (e.g., 512MB) in user-space, using a fixed mapping (e.g., 8 bytes of program memory to 1 byte of shadow state).
Such pre-allocation can perturb the target’s native memory layout and change subsequent allocation outcomes, which is particularly problematic for layout-sensitive workloads.
For instance, exploit code may rely on heap spraying to place objects in specific address ranges, while evasive malware may intentionally inflate code/data footprints.
In these cases, analysis-induced reservations can occupy address ranges needed by the workload, triggering allocation conflicts and causing divergence, crashes, or failure to reproduce the original behavior.

Second, to avoid in-process pollution, modern systems often decouple lightweight instrumentation from heavyweight tracking logic (e.g., running analysis in a separate process).
While decoupling improves transparency in principle, it introduces a new bottleneck: frequent cross-domain synchronization and IPC.
Fine-grained tracking generates high-rate events (e.g., memory accesses, taint propagation, control-flow metadata), and naïvely forwarding them through traditional IPC mechanisms amplifies overhead and can dominate runtime.
As a result, existing decoupled frameworks face a tension between maintaining high-fidelity tracking and keeping synchronization costs manageable, especially in interactive scenarios such as exploit reproduction and malware triage.

Therefore, we ask: can we build a fine-grained dynamic analysis framework that (1) preserves target address-space layouts to remain compatible with layout-sensitive exploits and evasive binaries, while (2) keeping decoupled tracking efficient by reducing synchronization/IPC overhead?

To answer these two questions, we propose  \textbf{H}ollowing \textbf{A}na\textbf{L}ysis \textbf{F}ramework (\textbf{HALF}), a fine-grained dynamic analysis framework for realistic Windows environments.
At its core, HALF follows a decoupled design that keeps lightweight instrumentation inside the target process while offloading heavyweight tracking logic to a separate analysis container; however, unlike prior decoupled systems, its key enabling principle is address-space transparency.
We leverage kernel-assisted process hollowing to construct a hollowed container that shares the same address layout with the target, thereby eliminating shadow-memory-induced layout perturbation and preventing allocation conflicts on layout-sensitive binaries.
To further mitigate the efficiency penalty of decoupling, we incorporate exception-driven synchronization, using shared memory together with page-fault/exception signals to coordinate cross-process state with minimal IPC traffic.
In doing so, HALF remains deployable on bare-metal Windows, avoids address-space pollution typical of in-process trackers, and reduces the synchronization overhead that often dominates fine-grained tracking.

We evaluate HALF on common applications, exploit binaries, and real-world malware, showing broad applicability and substantial performance gains.
Our main contributions are as follows:
\begin{itemize}
\item \textbf{Kernel-assisted Process Hollowing for Transparency.} We propose a kernel-assisted hollowing mechanism that constructs transparent, same-address analysis containers, eliminating address space conflicts for layout-sensitive binary analysis.
\item \textbf{Exception-driven Synchronization for Efficiency.} We design an exception-driven synchronization technique utilizing shared memory and page faults to minimize IPC bottlenecks inherent in decoupled analysis.
\item \textbf{Prototype, Evaluation, and Artifacts.} We implement a prototype on Windows and provide an artifact package with extensive evaluation, demonstrating the ability to analyze exploits/malware where traditional in-process baselines fail due to memory conflicts.
\end{itemize}

\section{Preliminaries}
\label{sec:prelim}

\subsection{Binary Program Analysis}
Binary program analysis underpins many security tasks, including exploit reproduction, malware understanding, and forensic reconstruction.
Compared to source-level analysis, binaries expose sparse semantic cues: type information and high-level boundaries are largely erased, and compiler optimizations and obfuscation further weaken structure recovery.
Consequently, analysts must reconstruct control-flow and data dependencies under incomplete information, and the analysis methodology directly shapes the fidelity and reproducibility of conclusions.

Static approaches operate without executing the target and typically rely on disassembly, control-flow recovery, and symbolic reasoning to infer behaviors and enumerate paths.
They scale well for broad coverage and are commonly used for large-scale scanning and feature extraction~\cite{zlin_2023_typesqueezer,mlim_2023_teasing}.
Static taint analysis has also been applied to large application ecosystems (e.g., mini-programs) to track sensitive flows at scale~\cite{cwang_2023_taintmini}.
However, binaries exhibit structural uncertainty (e.g., indirect branches and late-bound dispatch), and real-world samples frequently use packing, obfuscation, or self-modification, which break static assumptions and force conservative approximations~\cite{oor-meir_2019_dynamic,ebauman_2015_a,ddelia_2019_sok}.

Dynamic approaches observe real execution to reduce ambiguity and to capture runtime-only behaviors such as decrypted payloads and concrete allocation trajectories.
This practicality makes dynamic analysis a cornerstone for modern malware triage and sandboxing, but it also raises challenges in overhead, environment dependence, and deployability~\cite{akuchler_2021_does,ddelia_2019_sok,lchua_2019_one}.
Among fine-grained dynamic techniques, DTA tracks how input bytes propagate through registers and memory to form a causal evidence chain, which is central for explaining vulnerability triggers and reconstructing attack provenance~\cite{jschwartz_2010_all,klee_2013_high}.
Therefore, the core question becomes how to implement DTA with both high semantic fidelity and practical performance.

\subsection{DTA Implementation Mechanisms}
Fine-grained DTA can be realized through three mechanism families: hardware-assisted tracing, virtualization-based execution, and user-space instrumentation.
As summarized in Table~\ref{tab:dta-spectrum}, these approaches differ systematically along four dimensions: semantic visibility, runtime overhead, transparency, and deployability.
Their key differences, therefore, are not merely about whether they can execute the target, but which semantic granularity they can expose and what costs must be paid to obtain that granularity.

\begin{table}[!htbp]
\centering
\caption{A coarse comparison of DTA implementation mechanisms.}
\label{tab:dta-spectrum}
\footnotesize
\setlength{\tabcolsep}{4pt}
\renewcommand{\arraystretch}{1.15}
\renewcommand{\tabularxcolumn}[1]{m{#1}} % make X vertically centered

\begin{tabularx}{\linewidth}{|m{0.18\linewidth}|X|X|X|}
\hline
\multicolumn{1}{|c|}{\textbf{Dimension}} &
\multicolumn{1}{c|}{\textbf{Hardware tracing}} &
\multicolumn{1}{c|}{\textbf{Virtualization}} &
\multicolumn{1}{c|}{\textbf{Instrumentation}} \\
\hline
Semantic visibility &
Control-flow focused, limited operand values &
Whole-system, but process semantics require reconstruction &
Operand-level access in-process \\
\hline
Typical overhead &
Low for tracing, higher for data reconstruction &
High (Emulation/Virtual Machine (VM) tax) &
Moderate, often near-native for many workloads \\
\hline
Transparency &
Low perturbation, but limited semantics &
High isolation, strong against in-guest checks &
In-process footprint is often detectable \\
\hline
Deployability &
Requires hardware features/tooling &
Requires images/VM stack, heavier setup &
User-space deployment is flexible \\
\hline
\end{tabularx}
\end{table}

Hardware tracing (e.g., Intel Processor Trace (PT)-style designs) offers low-perturbation control-flow recording, enabling high-throughput collection of execution provenance~\cite{sma_2016_protracer}.
As Table~\ref{tab:dta-spectrum} indicates, its semantic visibility is primarily control-flow oriented, while operand values are not directly available.
Because DTA requires operand-level propagation evidence, data flow must be reconstructed through additional inference, which raises complexity and can reduce fidelity; the resulting cost often shifts from tracing overhead to reconstruction overhead.

VM/emulation-based analysis runs the target inside an isolated guest and keeps analysis logic outside the guest boundary, improving transparency against in-guest anti-analysis and enabling whole-system visibility~\cite{bdolan-gavitt_2015_repeatable,adavanian_2019_decaf,adinaburg_2008_ether}.
Table~\ref{tab:dta-spectrum} highlights two persistent frictions: high runtime overhead from virtualization/emulation, and a semantic gap where process-level objects must be recovered from low-level machine states.
This combination tends to increase engineering complexity and can amplify end-to-end analysis cost, especially when byte-level dependence tracking is required.
Pin-compatible out-of-VM DBI has also been explored to improve compatibility in virtualized setups~\cite{jzeng_2015_pemu}.

DBI frameworks such as Pin, DynamoRIO, and Valgrind (as well as lightweight engines such as TinyInst) rewrite instruction streams and execute instrumented basic blocks from a code cache, providing direct access to registers, memory operands, and per-instruction semantics~\cite{cluk_2005_pin,dbruening_2012_transparent,nnethercote_2007_valgrind,tinyinst_2023_tinyinst}.
Consistent with Table~\ref{tab:dta-spectrum}, instrumentation offers the strongest operand-level semantic visibility with practical deployability in user space, while its in-process footprint can reduce transparency against anti-instrumentation checks.
This makes instrumentation the dominant substrate for fine-grained DTA implementations such as libdft and TaintEraser~\cite{pkemerlis_2012_libdft,dzhu_2010_tainteraser}.
To further reduce overhead, prior work explores decoupling and parallelization (e.g., ShadowReplica, TaintPipe, StraightTaint, and PiTa), but still pays for high-rate synchronization and trace transfer~\cite{kjee_2013_shadowreplica,jming_2015_taintpipe,jming_2016_straighttaint,jgalea_2020_the}.
Other efforts reduce overhead through static rewriting and optimistic hybrid tracking~\cite{schen_2021_selectivetaint,sbanerjee_2019_iodine} or improve scalability via learned taint inference and GPU-assisted query representations~\cite{dshe_2020_neutaint,kji_2022_flowmatrix}.
Beyond native single-language binaries, taint analysis has been extended to cross-language and script-engine settings~\cite{wli_2022_polycruise,tusui_2022_script}, and to scenarios that require instrumentation across user-kernel boundaries or dynamic kernel analysis support~\cite{jhong_2021_a,ppitigalaarachchi_2023_krover}.
Finally, recent designs explore compact taint representations (e.g., container-tag-based tracking) to reduce metadata cost while preserving fine-grained semantics~\cite{zjia_2023_design}.
Alternative designs integrate tainting into managed runtimes or Operating System (OS) services (e.g., TaintDroid for Android), yet such designs are platform-specific and difficult to generalize to native Windows binaries~\cite{wenck_2010_taintdroid}.

Overall, when the analysis goal emphasizes byte-level propagation evidence, instrumentation provides the most direct semantic access with feasible throughput.
Thus, the main bottleneck shifts from whether to instrument to how to make instrumentation faithful for layout-sensitive binaries and efficient under decoupled tracking.

\subsection{Motivation: Address Space Conflicts}
\label{sec:motivation}

Instrumentation-based DTA typically shares the same address space between the target program and the analysis runtime, including the code cache and metadata.
In fine-grained tracking, large shadow states (such as tag maps and shadow memory) are often maintained within the target process itself, using fixed shadow layouts and large memory reservations~\cite{pkemerlis_2012_libdft,kserebryany_2012_addresssanitizer}.
This shared memory space can cause address conflicts, which may interfere with the target’s native memory layout and allocation patterns.
As a result, it can reduce the fidelity of the analysis, potentially disrupting the reproduction of vulnerabilities or breaking the evidence chain needed for tracing exploits.

\begin{figure}[!htbp]
\centering
\includegraphics[width=\linewidth]{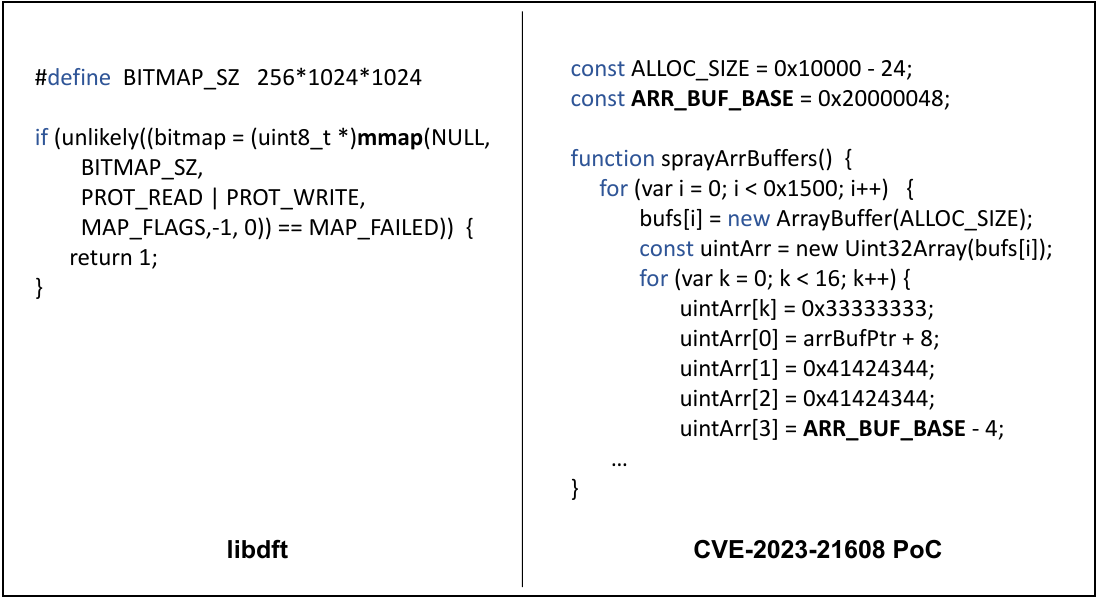}
\caption{A code-level example of an address-space conflict between a libdft-style in-process taint tracker and the layout-sensitive Adobe Reader PoC (CVE-2023-21608).
The tracker pre-reserves a large shadow region (left), while the PoC attempts to spray allocations into a fixed low-address range (right).
When these reservations overlap, the PoC cannot obtain the intended layout and may fail before triggering the vulnerability.}
\label{fig:libdft_code_conflict}
\end{figure}

This conflict becomes particularly acute for layout-sensitive workloads.
Many real-world applications rely on spatial determinism, meaning stable relative placements and allocation trajectories in memory.
For example, exploit techniques like heap spraying aim to place objects in specific address ranges.
However, if the analysis runtime reserves or occupies these ranges, the target’s allocator behavior can change, leading to deviations in the intended memory layout.
This can prevent the original behavior from being reproduced, cause crashes, or result in incomplete analysis.
In the case of libdft-style shadow reservations, conflicts can arise when heap spraying, as seen in an Adobe Reader Proof-of-Concept (PoC) (CVE-2023-21608), attempts to allocate in a specific range.
Fig.~\ref{fig:psmal} illustrates this issue at a high level, while Fig.~\ref{fig:libdft_code_conflict} provides a concrete code-level example of two competing allocation strategies.

Existing mitigations remain insufficient because they treat addressing conflicts as a minor deployment issue rather than a fundamental consequence of in-process analysis.
They try to negotiate space by shifting or reserving memory regions, assuming that the target can tolerate layout deformation.
However, the core problem remains: the analyzer and the target share the same virtual address space, causing them to compete for memory.
When DTA requires large, long-lived shadow reservations, any workaround just moves the footprint around, shrinking the target’s available memory space.
This issue becomes even more pronounced on 64-bit systems, where memory fragmentation and allocator non-determinism make “conflict avoidance” more difficult and still behavior-changing.

Addressing this fundamental issue requires removing the constraint of co-residency rather than negotiating for remaining space.
A practical approach is to create a separate process, clear its default user-space mappings, and arrange its memory exactly as required by the target before the analysis runtime allocates large regions.
This requirement motivates process hollowing, which provides control over the container’s initial address space layout.

\subsection{Process-level Isolation via Process Hollowing}

\label{sec:hollowing_bg}
To achieve high semantic fidelity while maintaining a clean, controlled environment, a system needs to be designed that minimizes address-space pollution and keeps analysis components completely separate from the target’s layout.

Process hollowing provides a concrete mechanism for isolating analysis components while preserving a target-like address-space layout.
It is commonly used as a technique for injection and evasion: a benign process is created in a suspended state, and its original image is replaced with a different payload that executes under the same process identity~\cite{mitre_attack_2025_process_hollowing,microsoft_2022_process_creation_properties}.
The motivation for attackers is stealth, since the payload inherits observable attributes of the original process (such as name and process identifier), making it harder to detect.
From an implementation perspective, hollowing transforms process creation into a programmable pipeline: it spawns a host process, identifies image metadata through the Process Environment Block (PEB), removes or neutralizes the original image, maps the replacement image by copying headers and sections, updates process bookkeeping, and rewrites the thread context to point to the new entry point before resuming execution.

In binary analysis, the value of process hollowing is the control it provides over the process creation phase rather than stealth.
By intercepting the process creation before execution begins, hollowing allows us to create a container process that perfectly matches the target’s native memory layout.
This enables us to keep analysis-only components (such as tracking code, shadow metadata, and buffers) outside the target’s memory space, linked only through controlled shared mappings.

This design requires a clear separation between the target’s memory layout, the container hosting the target, and the external analysis logic.
This architecture ensures that layout-sensitive exploits, such as heap spray attacks, encounter a memory environment that closely resembles a native execution context, effectively solving the issues posed by memory layout contamination and analysis overhead.

\section{System Design}
\label{sec:design}

Fine-grained dynamic analysis must balance two competing goals: (1) transparency, which requires preserving the target’s native address-space layout and execution behavior, and (2) efficiency, which requires keeping the hot path lightweight even when analysis maintains large, long-lived metadata.
Placing heavyweight tracking logic and shadow state inside the target often violates both goals by introducing large memory reservations and additional control flow.
HALF addresses this tension by decoupling execution from analysis: the target records a compact stream of dynamic facts, and a hollowed analysis container replays the trace under the same-address memory view reconstructed by a kernel monitor.

\begin{figure}[!htbp]
\centering
\includegraphics[width=\linewidth]{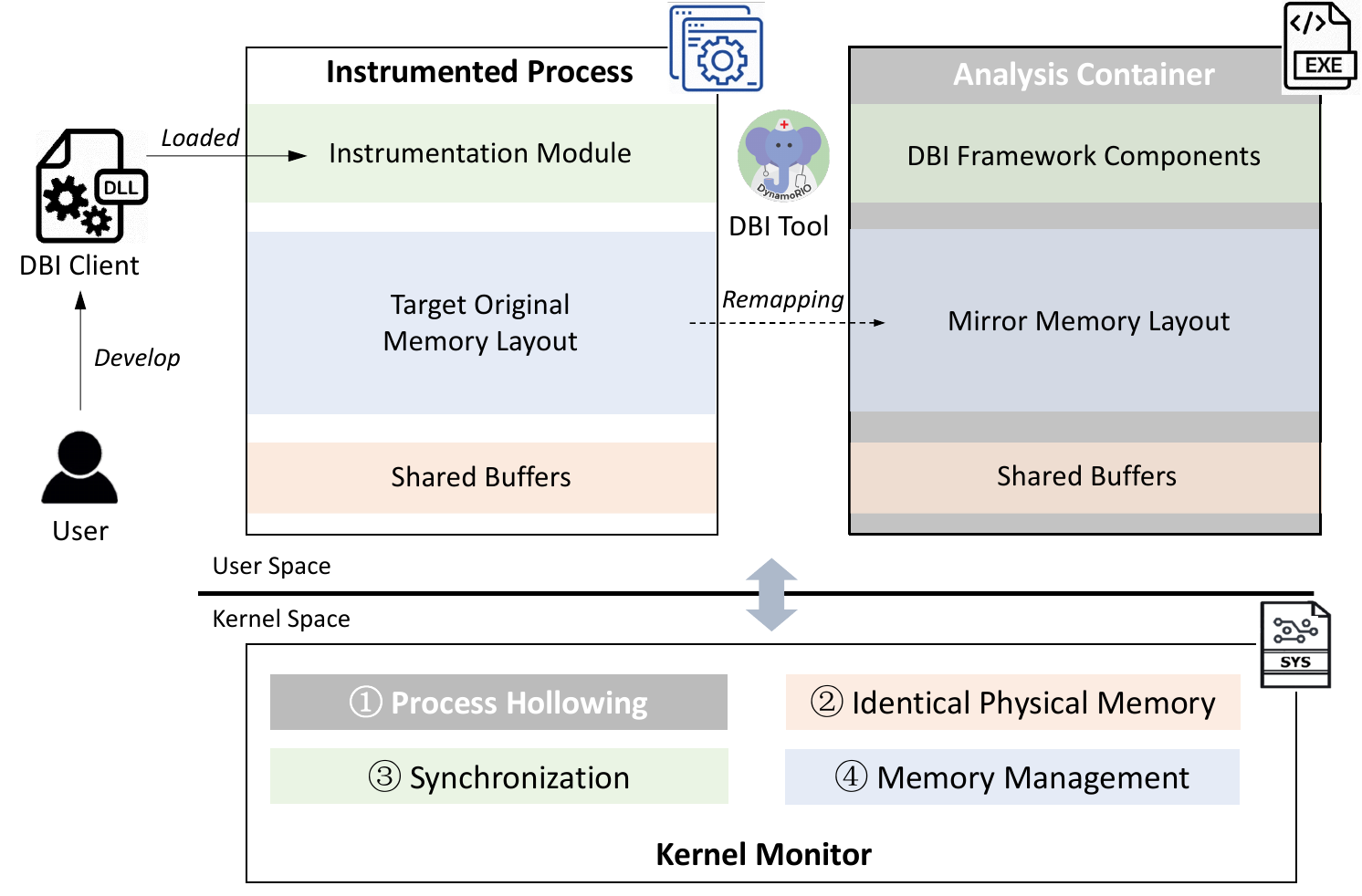}
\caption{HALF framework.
HALF consists of three cooperative entities: (1) an instrumented process hosting the recording logic, (2) an analysis container executing heavyweight tracking and task handlers, and (3) a kernel monitor coordinating memory reconstruction and synchronization.
The kernel monitor provides four mechanisms that jointly enforce transparency and efficiency: (1) process hollowing to construct a minimal container without inheriting the default user-space layout, (2) identical physical-memory mapping to share selected regions while keeping virtual addresses consistent, (3) exception-driven synchronization to move coordination off the hot path, and (4) on-demand memory management to materialize large analysis state without preemptive allocation conflicts.}
\label{fig:framework_half}
\end{figure}

Fig.~\ref{fig:framework_half} summarizes the architecture and assigns explicit roles to three entities: (1) the instrumented process (the target) acts as a lightweight data producer, (2) the analysis container (the container) provides an isomorphic execution environment for replay and analysis, and (3) the kernel monitor manages cross-process resources and mediates synchronization.
This separation relies on four kernel mechanisms: (1) process hollowing to construct a minimal container while preserving the target view, (2) identical physical-memory mapping to share selected regions under consistent virtual addresses, (3) exception-driven synchronization to offload buffer switching and coordination from the hot path, and (4) on-demand memory management to commit analysis-state pages only when accessed.
With these mechanisms, each target thread ((T_i)) is paired with a container thread ((A_i)) that consumes (T_i)'s record buffers and executes tracking logic asynchronously, enabling fine-grained analysis while maintaining a near-native target memory layout.

\subsection{Instrumentation Module: Minimalist Recording and Offloading}
\label{sec:instrumentation}
Target-side instrumentation must avoid address-space pollution and minimize hot-path control flow, because large reservations and frequent checks can perturb layout-sensitive programs and inflate execution overhead.
To meet this requirement, HALF implements instrumentation as a DBI client loaded into the target and restricted to DBI-provided interfaces.
It keeps the deployment model stable by avoiding modifications to the DBI engine source code or binaries, and it confines target-side logic to recording and code generation while delegating synchronization and privileged operations to the kernel monitor.

For fine-grained tracking, the client records only dynamic facts that are expensive to recover statically, including effective addresses for memory operands, selected bits of the EFLAGS register, and dynamic operands such as \texttt{cl} used by shift instructions.
It keeps the original instruction stream intact and injects a short straight-line stub that captures operands primarily via simple instructions such as \texttt{lea} and \texttt{mov}.
For complex instructions (e.g., \texttt{rep movs}), it records dependent registers instead.
Recording state (e.g., record-buffer pointers and temporary register-save slots) is stored in per-thread locations reachable through GS/TEB.

\begin{figure}[!htbp]
\centering
\includegraphics[width=0.9\linewidth]{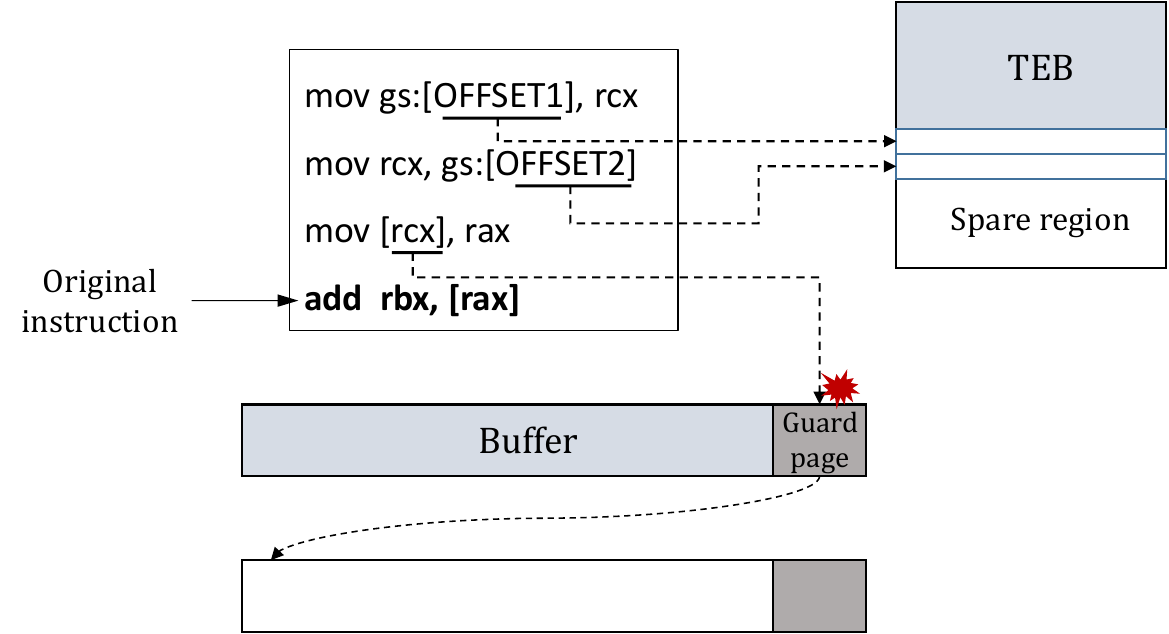}
\caption{Runtime information recording optimization scheme diagram.
(1) Fast record-pointer access via GS/TEB slots.
(2) Zero-check buffer switching via guard-page faults: touching the guard page transfers control to the kernel monitor, which rotates the writer to a new free buffer.}
\label{fig:recordopt}
\end{figure}

Conventional recorders perform explicit bounds checks (or branch-based fullness tests) before appending each record, which introduces frequent control flow on the hot path.
HALF instead adopts a zero-check design: each buffer ends with a non-writable guard page, so an overflow write triggers a Page Fault (\#PF) that the kernel monitor handles by rotating the writer to a fresh buffer (Fig.~\ref{fig:recordopt}).
This shifts buffer switching to rare exceptions and eliminates per-record checking overhead.

To avoid replaying raw instruction semantics in the container, the client generates a native analysis block for each instrumented basic block and places it into executable pages that will be shared with the container.
It records only the entry address of each generated block into the record stream, enabling the container to replay the trace by direct jumps.
To support analyst-defined checks without inflating instruction-level propagation logic, the client can also emit short task-invocation stubs at selected program points.
Each stub encodes a task ID and arguments and then transfers control to a container-side dispatcher (Fig.~\ref{fig:analysistask}), while task logic is implemented in the container (Section~\ref{sec:container}).

This design limits the target hot path to straight-line recording and occasional exception transitions at buffer boundaries, while heavyweight tracking and task handlers run in the container under the same-address memory view.
Because the record stream is intentionally compact and does not carry semantic context or address-translation metadata, correct interpretation of recorded pointers requires an execution environment that preserves same-address semantics.
This motivates the hollowed analysis container.

\subsection{Analysis Container: Hollowed Process for Isomorphic Replay}
\label{sec:container}
Heavyweight tracking logic and large shadow state are difficult to host inside the target without perturbing its layout.
Moving analysis out of process avoids target-side footprint, but it must still preserve pointer semantics so that recorded addresses can be dereferenced consistently.
For this reason, HALF constructs a hollowed container and remaps selected regions to match the target at the same virtual addresses, so recorded pointer values remain directly valid.
In particular, DBI-reserved ranges, shared record buffers, and executable analysis-code pages are reconstructed in the container with the same address map.
For shared regions, the kernel monitor can also map identical physical pages into both processes.
This design removes the need for address translation in the analysis hot path and keeps pointer-based semantics consistent across the target and the container.
To reduce interference from user-mode runtime components, the container retains only a minimal runtime after hollowing and accesses required OS services via thin wrappers or direct system calls.

\begin{figure}[!htbp]
\centering
\includegraphics[width=0.9\linewidth]{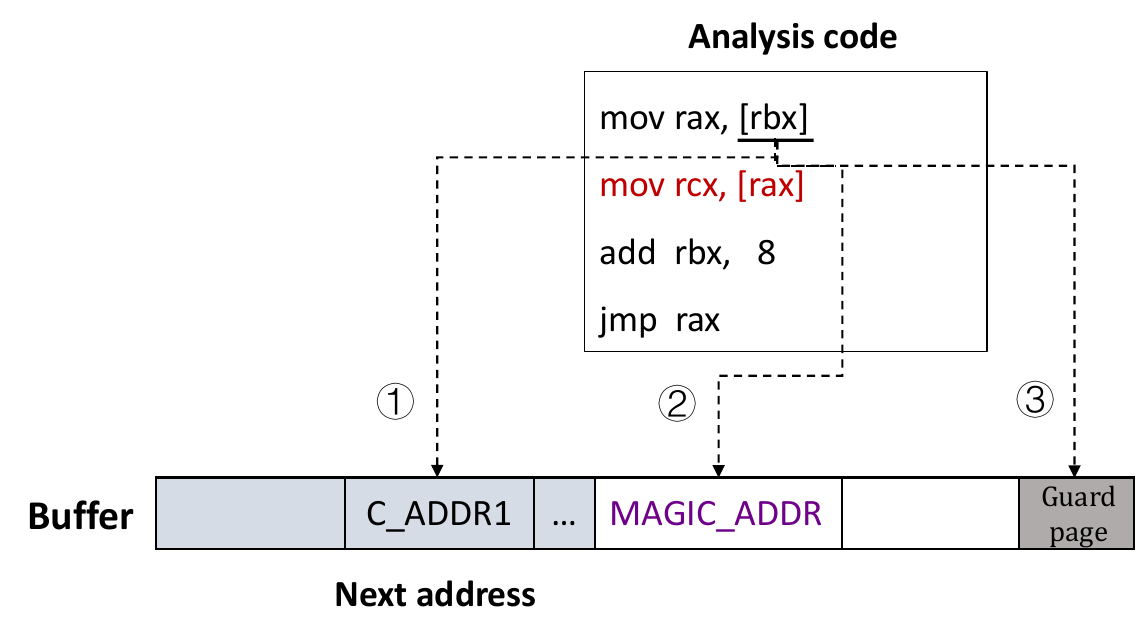}
\caption{Analysis code execution switching.
(1) Normal read: load a valid code address and \texttt{jmp}.
(2) Sentinel switching: \texttt{MAGIC_ADDR} triggers an exception used by the kernel monitor to rotate buffers.
(3) Guard-page switching: touching the buffer-end guard page triggers a page fault for rotation.}
\label{fig:aces}
\end{figure}

During execution, HALF binds a corresponding analysis thread ((A_i)) in the container for each target thread ((T_i)).
(A_i) consumes (T_i)'s record buffers and replays control flow by reading the next analysis-block entry address from the record stream and jumping directly to it; buffer boundaries and mid-buffer rotations are handled by exceptions rather than explicit checks (Fig.~\ref{fig:aces}).

\begin{figure}[!htbp]
\centering
\includegraphics[width=\linewidth]{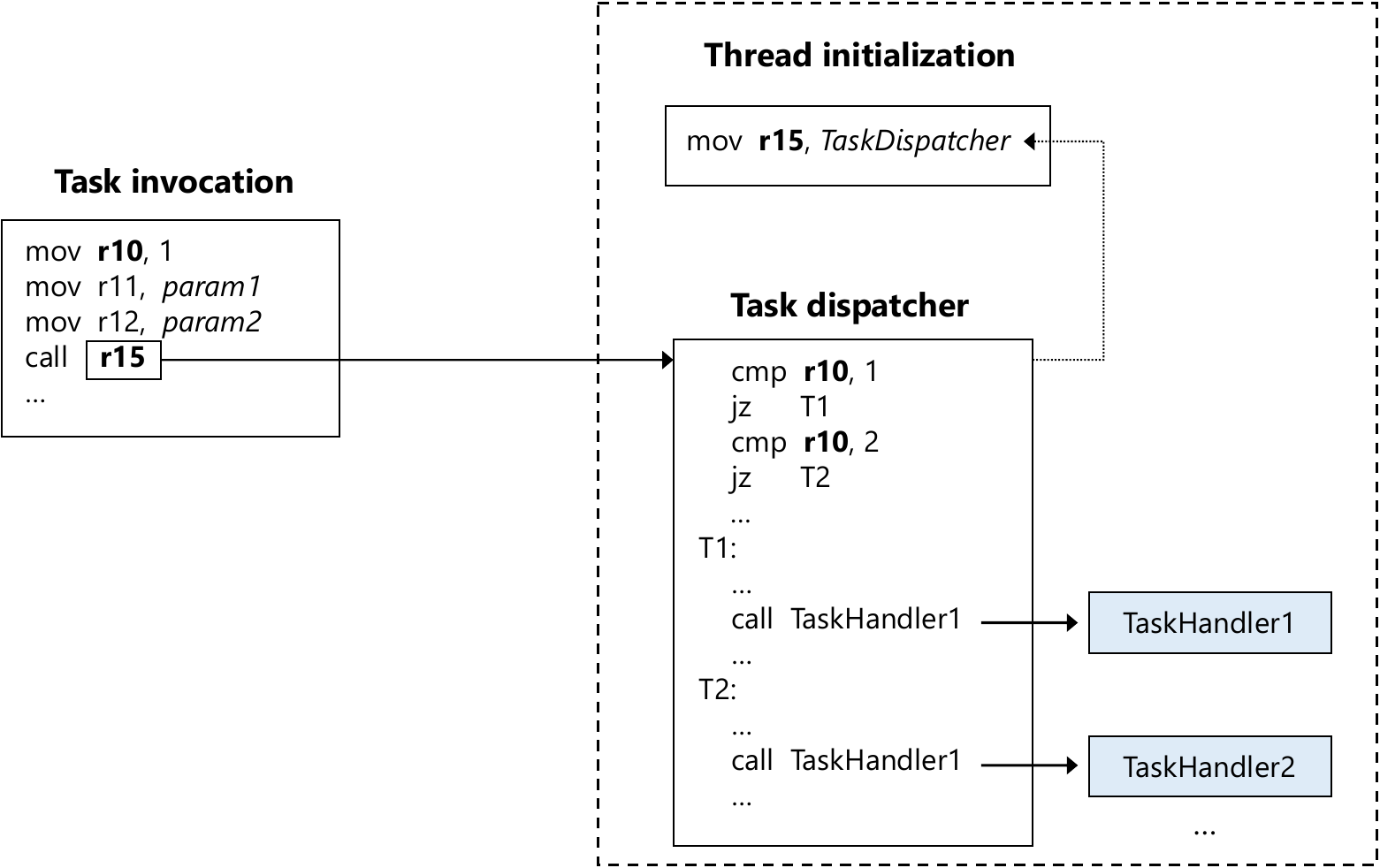}
\caption{Analysis task invocation and dispatch.
The target-side stub loads the task ID (e.g., into \texttt{r10}) and calls a dispatcher entry (e.g., via \texttt{r15}).
The dispatcher branches on the task ID and invokes the corresponding handler implemented in the container.}
\label{fig:analysistask}
\end{figure}

To express analysis goals such as sources/sinks, API-level invariants, and reporting actions, HALF supports tasks that run as container-side handlers.
To invoke a task, the target-side stub loads a task ID and arguments according to a fixed convention and transfers control to a dispatcher entry point in the container (Fig.~\ref{fig:analysistask}).
To keep the stub compact and stable across tasks, the dispatcher address is prepared when the analysis thread starts.
In our prototype, \texttt{r10} carries the task ID and \texttt{r15} points to the dispatcher entry, while the remaining arguments use designated registers or stack slots depending on the task.

To keep the container robust after hollowing, task handlers avoid relying on conventional DLL loading and instead use container-provided wrappers for common services.
To perform privileged operations when needed, handlers can request assistance from the kernel monitor through dedicated control interfaces.

As a result, the container can dereference recorded pointers directly and execute heavyweight tracking in parallel with target execution, while the target remains largely unmodified and layout-stable.
Maintaining the isomorphic view and mediating exception-driven replay require privileged address-space control and cross-process exception handling, which are provided by the kernel monitor.

\subsection{Kernel Monitor: Resource Coordination and Exception Mediation}
\label{sec:kernel}
Preserving same-address semantics and coordinating decoupled execution require privileged control over address spaces and exception paths that is not reliably available in user mode, especially when the DBI engine must remain unmodified.
HALF therefore implements a kernel monitor as a Windows kernel driver that orchestrates container hollowing, address-space reconstruction, and cross-process synchronization.
It observes target allocations to infer DBI-reserved ranges, replays the same reservations in the container (often as reserve-without-commit), and uses page-table operations to map shared regions at consistent virtual addresses across the target and container.
During execution, it mediates page-fault and sentinel-triggered exceptions for buffer rotation and handles selected synchronization-related syscalls to bound analysis lag.
This centralizes resource management and synchronization decisions in a privileged component, keeping the target-side instrumentation lightweight while preserving same-address transparency.

\begin{figure*}[!t]
\centering
\includegraphics[width=0.95\textwidth]{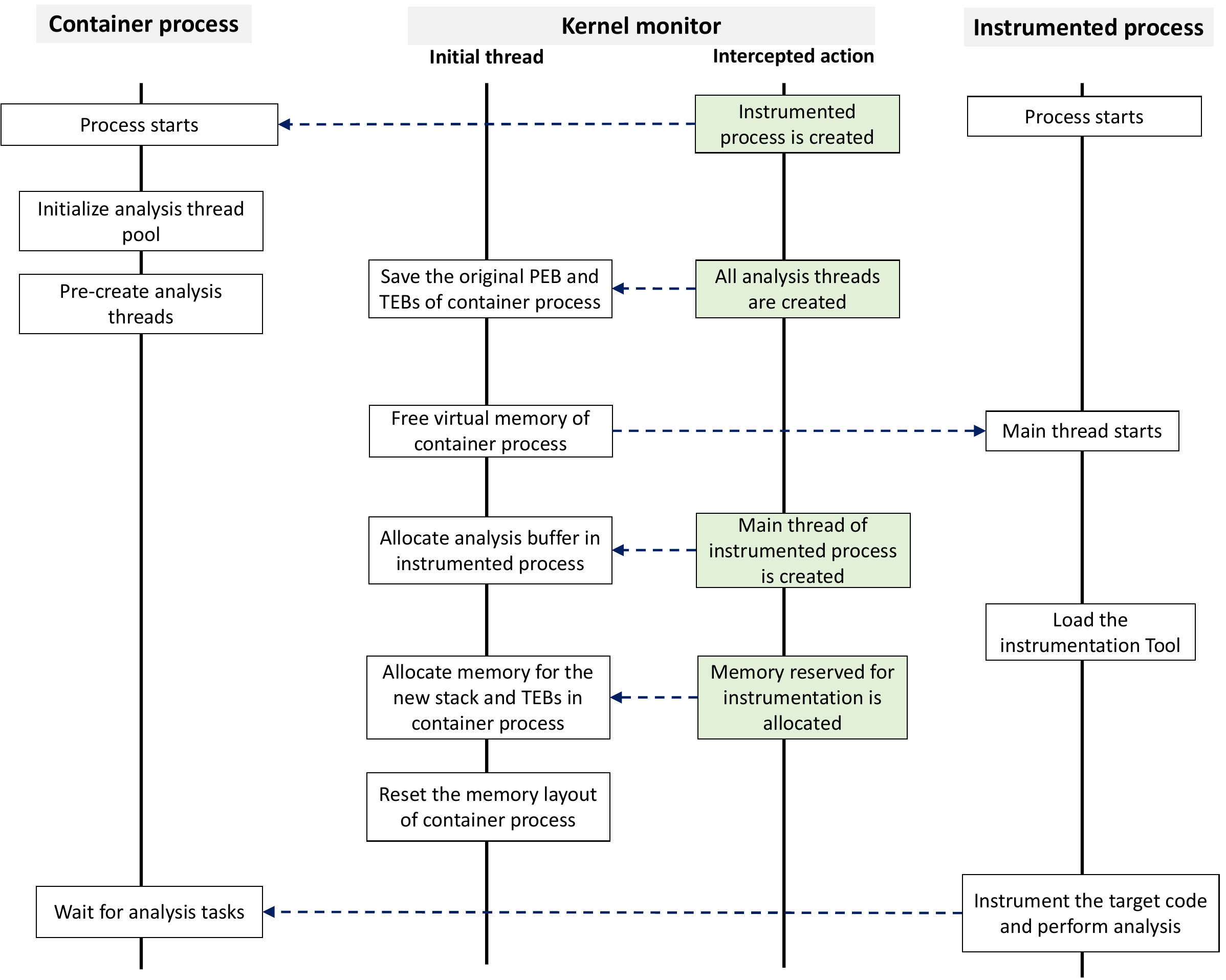}
\caption{\normalsize Initialization sequence diagram.
The kernel monitor hollows the container, reconstructs a minimal runtime, and remaps key regions (shared buffers and DBI-reserved ranges) to align with the target view.
It also rebuilds the analysis-thread runtime state (TEB/stack) before analysis begins.}
\label{fig:initpro}
\end{figure*}

\begin{figure}[!hbtp]
\centering
\includegraphics[width=\linewidth]{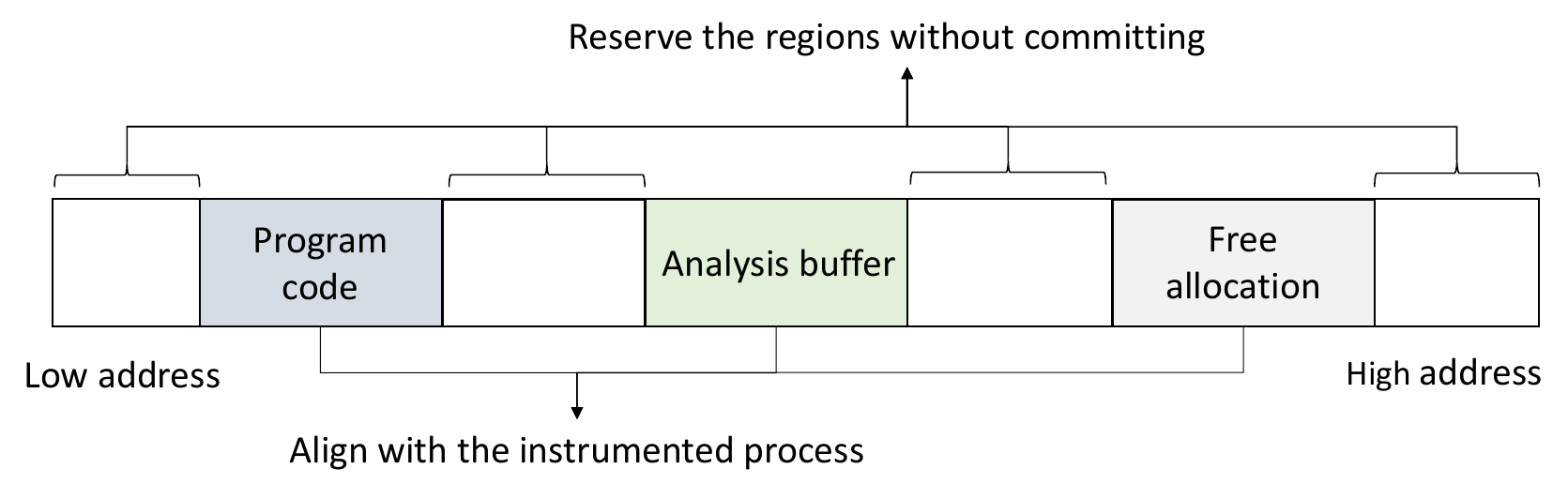}
\caption{\normalsize Container address-space layout after initialization.
Framework code regions, shared record buffers, and DBI-reserved ranges match the target layout.
Reserve-without-commit preserves the address map while reducing physical memory pressure; a free region is used for analysis-thread TEBs and stacks.}
\label{fig:memlayout}
\end{figure}

Before instrumented code runs, the monitor starts from a normal container process and hollows it by unmapping most user-space regions, keeping only the minimal code and data required for analysis.
It traverses container mappings, unmaps loaded modules, and frees committed regions (including PEB/TEB and stacks after adjusting protections), leaving only the startup code region that hosts the minimal runtime.
It then mirrors the target layout for DBI-reserved ranges and shared regions (Figs.~\ref{fig:initpro} and~\ref{fig:memlayout}), and coordinates multi-stage initialization across the target, the container, and the driver via kernel synchronization objects and event-driven callbacks.

\begin{figure}[!htbp]
\centering
\includegraphics[width=0.9\linewidth]{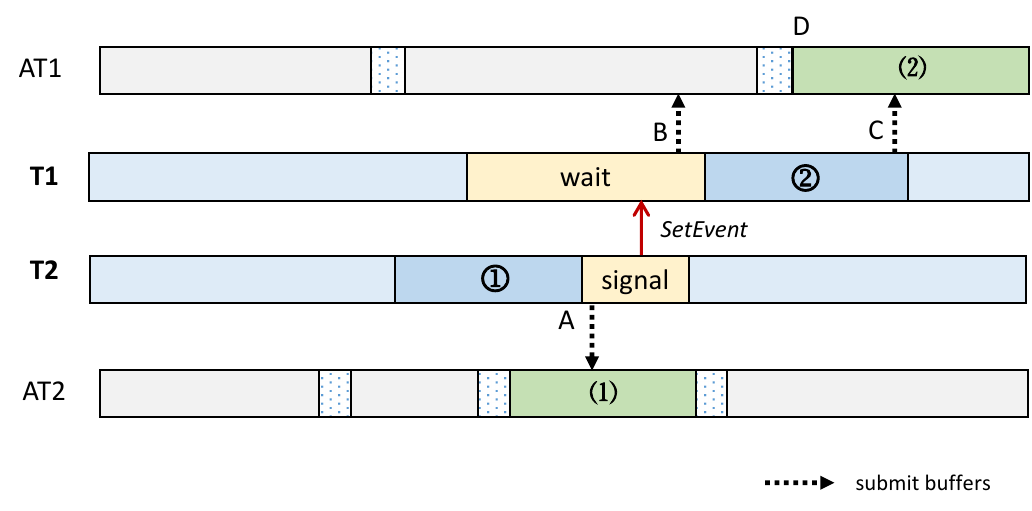}
\caption{Code synchronization example.
When a thread signals or returns from a wait, the kernel monitor forces the current record buffer to be committed so the corresponding analysis thread can process relevant records promptly, reducing under-analysis at synchronization edges.}
\label{fig:atsync}
\end{figure}

To avoid frequent polling checks in both producer and consumer, HALF uses exception paths for buffer management.
On the write side, a non-writable guard page at the end of the current record buffer triggers a page fault (\#PF), the monitor selects a fresh buffer from the Free list, and updates producer pointers.
On the read side, either reaching the buffer-end guard page or dereferencing a sentinel value (e.g., \texttt{MAGIC_ADDR}) triggers an exception that the monitor uses to switch the consumer to the next committed buffer (Figs.~\ref{fig:recordopt} and~\ref{fig:aces}).
To reduce analysis lag in multi-threaded targets, the monitor filters synchronization-related syscalls such as \texttt{NtWaitForSingleObject} and \texttt{NtSetEvent} and forces an early commit of the current record buffer even if it is not full (Fig.~\ref{fig:atsync}).
HALF does not attempt to reconstruct a complete happens-before graph; instead, it uses syscall-filtered early commits as a deployable approximation, and remaining limitations are discussed in Section~\ref{sec:limitations}.

These kernel mechanisms depend on a unified mapping strategy for shared regions and for the large shadow state, which we describe next.

\subsection{Memory Management and Cross-Component Communication}
\label{sec:memory_communication}
\begin{figure}[!htbp]
\centering
\includegraphics[width=\linewidth]{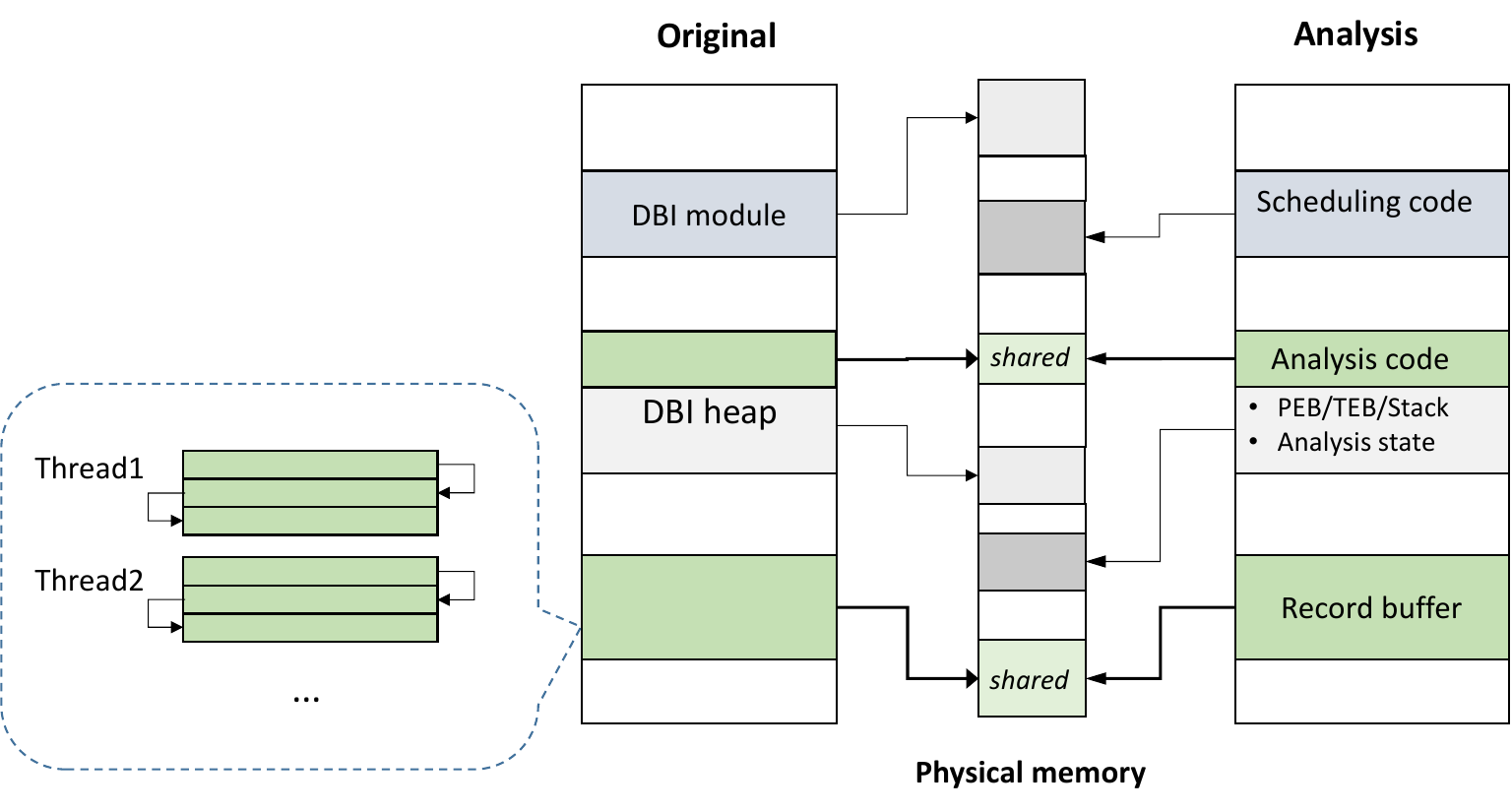}
\caption{Memory allocation and mapping.
The kernel monitor maps three regions into both processes at identical virtual addresses: scheduling/metadata stubs, executable analysis-code pages generated by the DBI client, and per-thread circular record buffers for streaming execution records.}
\label{fig:memalloc}
\end{figure}

To avoid collisions with application allocations while preserving pointer semantics, the kernel monitor places framework-managed shared regions inside DBI-reserved address ranges rather than the target’s normal allocation ranges.
These regions are mapped at the same virtual addresses in both the target and the container, and selected regions can be backed by identical physical pages to enable zero-copy sharing (Fig.~\ref{fig:memalloc}).
This layout allows both processes to directly load/store shared metadata and jump to shared analysis code without a translation layer.

The driver allocates a per-thread set of record buffers and maps them RW into both processes.
To enable continuous producer/consumer overlap, each thread uses multiple buffers arranged as a circular list with Free and Full lists: the target thread dequeues from Free, appends records, and commits to Full, the analysis thread dequeues from Full and recycles buffers back to Free.
To reduce address-space fragmentation, the driver can reserve record-buffer virtual address space early and assign fixed-size slices to threads as they are created.
Executable analysis-code pages generated by the DBI client and small stub/metadata pages (e.g., per-thread \texttt{Base}/\texttt{Limit}/\texttt{Current} pointers and status flags) are shared using the same mapping strategy.

\begin{figure}[!htbp]
\centering
\includegraphics[width=\linewidth]{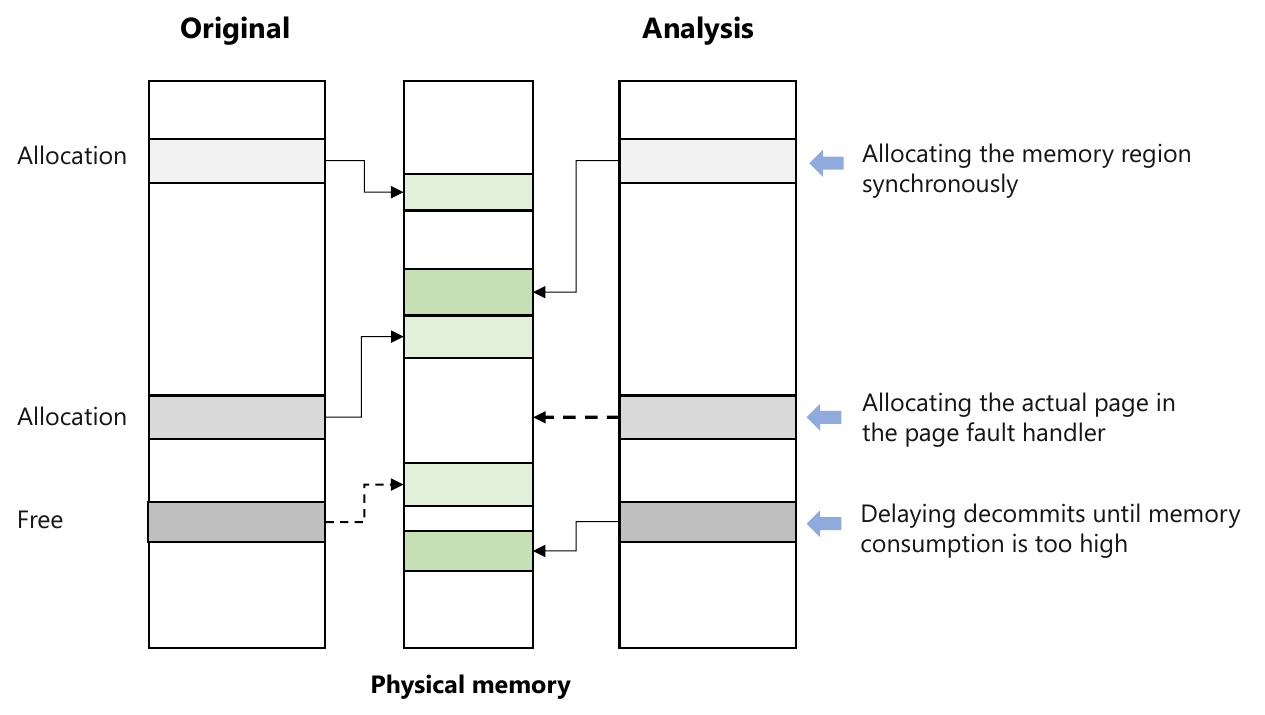}
\caption{Shadow memory allocation and management scheme.
Explicit target commits are mirrored to the container, container faults trigger on-demand commits, and frees can be delayed to reduce churn while preserving virtual layout stability.}
\label{fig:shadowmem}
\end{figure}

To support large analysis states without perturbing the target layout, HALF materializes shadow pages in the container on demand (Fig.~\ref{fig:shadowmem}).
To keep memory pressure low in typical workloads, large virtual reservations can exist without committing physical pages, since actual application footprints are often much smaller than the address space they reserve.
On the explicit path, the kernel monitor intercepts target memory commits through kernel hooks and commits corresponding container pages.
On the implicit path, when analysis code touches an uncommitted shadow page, a container-side page fault (\#PF) occurs, the monitor intercepts the exception, commits the page, and can trigger a kernel-mode touch to ensure a physical page is allocated before returning to user mode.
To reduce churn on memory release, the container does not immediately free the corresponding virtual pages when the target frees memory.
Instead, physical reclamation can be delayed; if memory consumption becomes too large, the monitor can evict some physical pages and temporarily spill their contents to disk, then reload them when future analysis accesses trigger page-fault exceptions.

The target and container use driver-control interfaces (e.g., IOCTLs) for one-time initialization and per-thread registration.
During steady state, exception paths drive buffer switching and early commits: write-side guard-page faults rotate producer buffers, and sentinel/guard exceptions rotate consumer buffers and can trigger commit at synchronization points.
For lightweight status exchange, shared metadata can also include simple atomic event flags to communicate thread states (e.g., wait/wake) between the target and the container.
The resulting layout enables zero-copy sharing of code and records, preserves same-address pointer semantics, and avoids large target-side reservations by committing analysis state on demand.

\section{Experiments and Analysis}
We implemented HALF on Windows using DynamoRIO (\texttt{cronbuild-9.91.19426}) as the instrumentation substrate.
The prototype is built upon three distinct parts:
a kernel monitor driver, a hollowed container process to mirror the target layout, and a DynamoRIO client DLL for instrumentation and DTA.
By confining every change to these components, we ensure the DynamoRIO core remains untouched.
Therefore, the proposed mechanisms are non-invasive and eliminate the need for forking industrial DBI stacks.

Our evaluation is structured to answer three design-aligned questions:
\begin{itemize}
\item \textbf{(Q1) Efficiency:} Does zero-check recording and exception-driven buffer switching remove hot-path branch costs and keep recording overhead low?

\item \textbf{(Q2) Transparency:} Does kernel-assisted process hollowing eliminate address-space interference on layout-sensitive binaries?

\item \textbf{(Q3) Practicality:} Can decouple, same-address analysis sustain real workflows (multi-module applications, exploits, and malware) with high semantic fidelity?

\end{itemize}
All experiments were conducted on an Intel i5-12400 2.50GHz (12 cores) Central Processing Unit (CPU) with 16GB memory, running Windows 10 (20H2) x64.
An anonymized artifact package (driver, container, instrumentation DLL, build scripts, and test cases) is available at \url{https://github.com/xxx/xxxx}.

To facilitate reproduction, we provide the following configuration details:
\begin{itemize}
\item \textbf{OS Version:} Windows 10 20H2 / 21H1 (x64).
Driver signing requires test-signing mode enabled (\texttt{bcdedit /set testsigning on}).
\item \textbf{DBI Platform:} DynamoRIO v9.91 or newer.
\item \textbf{Compilation:} Visual Studio 2019 + Windows Driver Kit (WDK) 10.0.19041.0.
\item \textbf{Benchmarks:} SPEC CPU2017 (Windows), plus real applications, exploit PoCs, and real-world malware, all compiled/executed on Windows where applicable.
\item \textbf{Execution:} Run via the provided PowerShell script: \texttt{PowerShell.exe -File run_analysis.ps1 -Target <input.exe>}.
\end{itemize}

\subsection{Implementation and Capability Comparison}
Our prototype instantiates the design in Section~\ref{sec:design}: (i) instrumentation remains lightweight in the target, (ii) heavyweight taint propagation runs in the container, and (iii) kernel-assisted process hollowing and same-address mappings prevent address-space pollution.

\begin{table}[!htbp]
\centering
\caption{Feature comparison with state-of-the-art binary analysis frameworks.}
\label{table:capability}
\footnotesize
\setlength{\tabcolsep}{3pt}

\begin{tabularx}{\linewidth}{|>{\centering\arraybackslash}X|c|c|c|c|}
\hline
\textbf{Feature} &
\textbf{Pin/DynamoRIO (DR)} &
\textbf{Libdft} &
\textbf{PANDA} &
\textbf{HALF} \\
& \textbf{(DBI Base)} & & & \\
\hline
Low Address-Conflict Risk & \Checkmark & \XSolidBrush & \Checkmark & \Checkmark \\
Same-Address View & \textemdash & \XSolidBrush & \XSolidBrush & \Checkmark \\
Kernel-Assisted Remap & \XSolidBrush & \XSolidBrush & N/A & \Checkmark \\
Layout-Sensitive Compatibility* & \Checkmark & \XSolidBrush & \Checkmark & \Checkmark \\
Native Performance & \Checkmark & \Checkmark & \XSolidBrush & \Checkmark \\
\hline
\multicolumn{5}{p{0.9\linewidth}}{\scriptsize{* Ability to execute layout-sensitive exploit PoCs without analysis-induced layout conflicts (e.g., heap spraying).}} \\
\end{tabularx}
\end{table}

Table~\ref{table:capability} positions HALF against representative alternatives.
We include Pin/DynamoRIO as a DBI-based reference point: a bare DBI engine often avoids massive in-process shadow reservations, but it does not provide instruction-grain taint semantics without adding substantial metadata and propagation logic.
In-process taint trackers (e.g., \textit{libdft}) provide fine-grained semantics but must co-reside with the target and therefore compete for virtual address space when large, long-lived shadow states are required.
VM-based platforms (e.g., PANDA) isolate analysis logic but typically sacrifice native performance and still cannot offer a same-address, pointer-preserving view to an external user-space analyzer.
HALF combines these properties: hollowing creates an analysis container with a target-matching layout, and the same-address view enables the analyzer to interpret target pointers directly, while large shadow states are mapped on-demand in the container rather than reserved in the target.

\subsection{Performance Analysis}
We rely on SPEC CPU2017 to stress the instrumentation pipeline.
Every benchmark was built with Visual Studio 2019 for Windows, and we ran the \textit{ref} or \textit{train} workloads so each case lasts tens of seconds for a stable signal.
Each experiment was repeated 10 times, and we report average values.
Because SPEC CPU2017 has limited Windows/MSVC compatibility, we report results on the subset we can build and run reliably on Windows: 505.mcf_r, 508.namd_r, 519.lbm_r, 520.omnetpp_r, 523.xalancbmk_r, 525.x264_r, 526.blender_r, 531.deepsjeng_r, 538.imagick_r, 541.leela_r, 544.nab_r, and 557.xz_r.
Our goal is not just to list slowdowns, but to tie them to concrete mechanisms: zero-check recording (using rare exceptions instead of per-instruction bounds tests), exception-driven decoupling (fault-driven buffer/commit coordination), and containerized shadow state (mapping metadata outside the target while keeping a same-address view).

\subsubsection{Data Recording Efficiency}
We first isolate the overhead of runtime data recording, which is a necessary substrate for decoupled DTA.
As a user-space baseline, we use DynamoRIO’s \textit{memtrace_x86} tool, slightly modified to record only the memory-address stream required by HALF.
Unless stated otherwise, HALF uses a 512-kilobyte (KB) per-thread record buffer (our default).
For this recording-only microbenchmark, we intentionally use a smaller 64 KB buffer and immediately recycle buffers upon filling (no disk I/O) to stress buffer switching and expose the cost of boundary management.

\begin{figure}[!htbp]
\centering
\includegraphics[width=\linewidth]{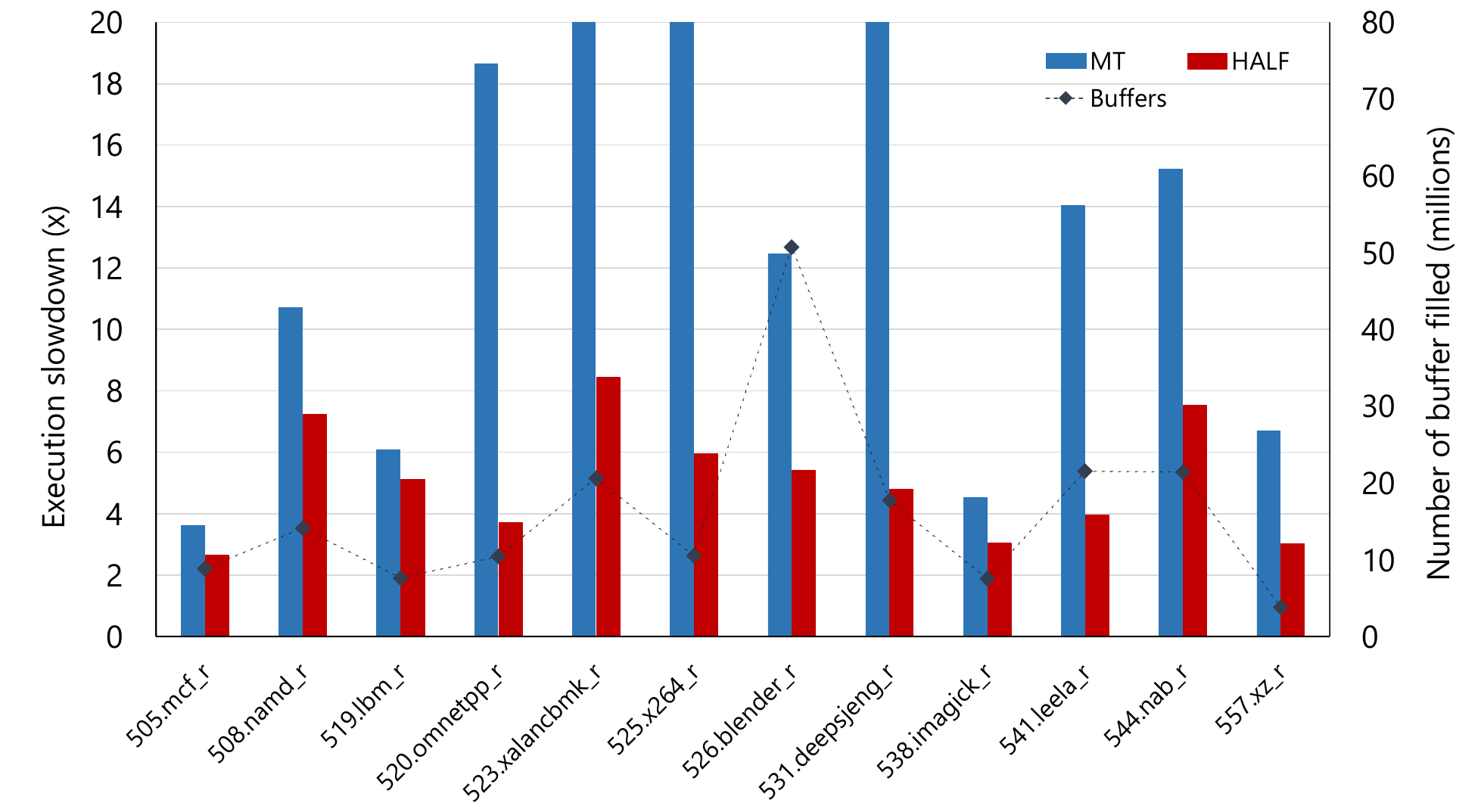}
\caption{Runtime overhead of memory address recording (HALF vs.
DynamoRIO memtrace).
The blue bars represent the slowdown of the baseline memtrace (MT) tool, while the red bars show HALF’s recording slowdown (left Y-axis).
The dashed line tracks the number of record buffer fills (right Y-axis, in millions).
HALF consistently outperforms the user-space recorder, reducing overhead by up to 15$\times$ on memory-intensive tasks (e.g., 525.x264_r, 531.deepsjeng_r), demonstrating the benefits of kernel-offloaded buffer management.}
\label{fig:perfover1}
\end{figure}

As shown in Fig.~\ref{fig:perfover1}, HALF substantially reduces recording overhead compared to the user-space baseline, especially for memory-intensive workloads.
Across most programs, HALF improves throughput by 2$\times$–8$\times$, for \textit{x264} and \textit{deepsjeng}, the reduction exceeds 15$\times$.
This is not an empirical coincidence but a direct consequence of the \textbf{zero-check mechanism}: rather than inserting a conditional branch on every recorded event to test buffer capacity, HALF uses a guard page so that a buffer boundary triggers a hardware exception and the kernel monitor rotates buffers.
This shifts the dominant cost from per-event branching to rare exception handling, while also reducing instruction bloat in hot basic blocks.
In addition, we leverage kernel-assisted and thread-local-storage-assisted context access and merge pointer updates at the basic-block level to minimize metadata traffic during steady state.

From a micro-architectural perspective, the performance gain on compute-intensive benchmarks like \textit{x264} can be attributed to improved \textbf{instruction locality}.
Traditional instrumentation injects branching logic (e.g., \texttt{cmp/jcc}) into every basic block to manage buffer capacity, which not only bloats code size but also increases pressure on the L1 instruction cache and consumes Branch Target Buffer (BTB) entries.
By offloading boundary checks to the Memory Management Unit (MMU) via guard pages, HALF preserves the compactness of the target’s hot loops.
This allows the CPU to sustain high instruction retirement rates, effectively decoupling the \textit{cost of observation} from the \textit{complexity of the workload}.

Fig.~\ref{fig:perfover1} also plots the frequency of buffer fills (dashed line).
Even with a 64KB buffer (a worst-case setting), buffer fills occur at the million scale, yet they do not dominate end-to-end runtime.
The experimental results demonstrate a clear separation between micro-level events and macro-level execution.
Specifically, the millions of exceptions are dwarfed by the massive volume of dynamic instruction and memory events in SPEC.
This distinction validates the arithmetic behind our decision to use exceptions instead of branches.
Since the contribution of buffer-switch faults is negligible, we conclude that hardware exceptions successfully amortize boundary management overhead and eliminate the instruction-level branching costs inherent in traditional instrumentation.
This also explains why the largest gains (e.g., $>$15$\times$) are observed on memory-intensive workloads: removing per-event checks eliminates a systematic cost that scales with the number of recorded events, while exception handling remains amortized.
However, this design implies a boundary: if a workload forces pathological buffer turnover (e.g., extremely small buffers under very high memory-event rates or unusually expensive exception dispatch due to third-party kernel instrumentation), exception handling can become visible in the profile.
Our default configuration (512KB per-thread buffers) is chosen to keep buffer switching safe in the amortized regime for typical workloads.

\subsubsection{End-to-End Analysis Overhead}
Here ``End-to-End’’ refers to measuring the complete workflow from target-side recording through container-side propagation, so the reported overhead reflects the full pipeline rather than a single component.
\begin{figure}[!htbp]
\centering
\includegraphics[width=\linewidth]{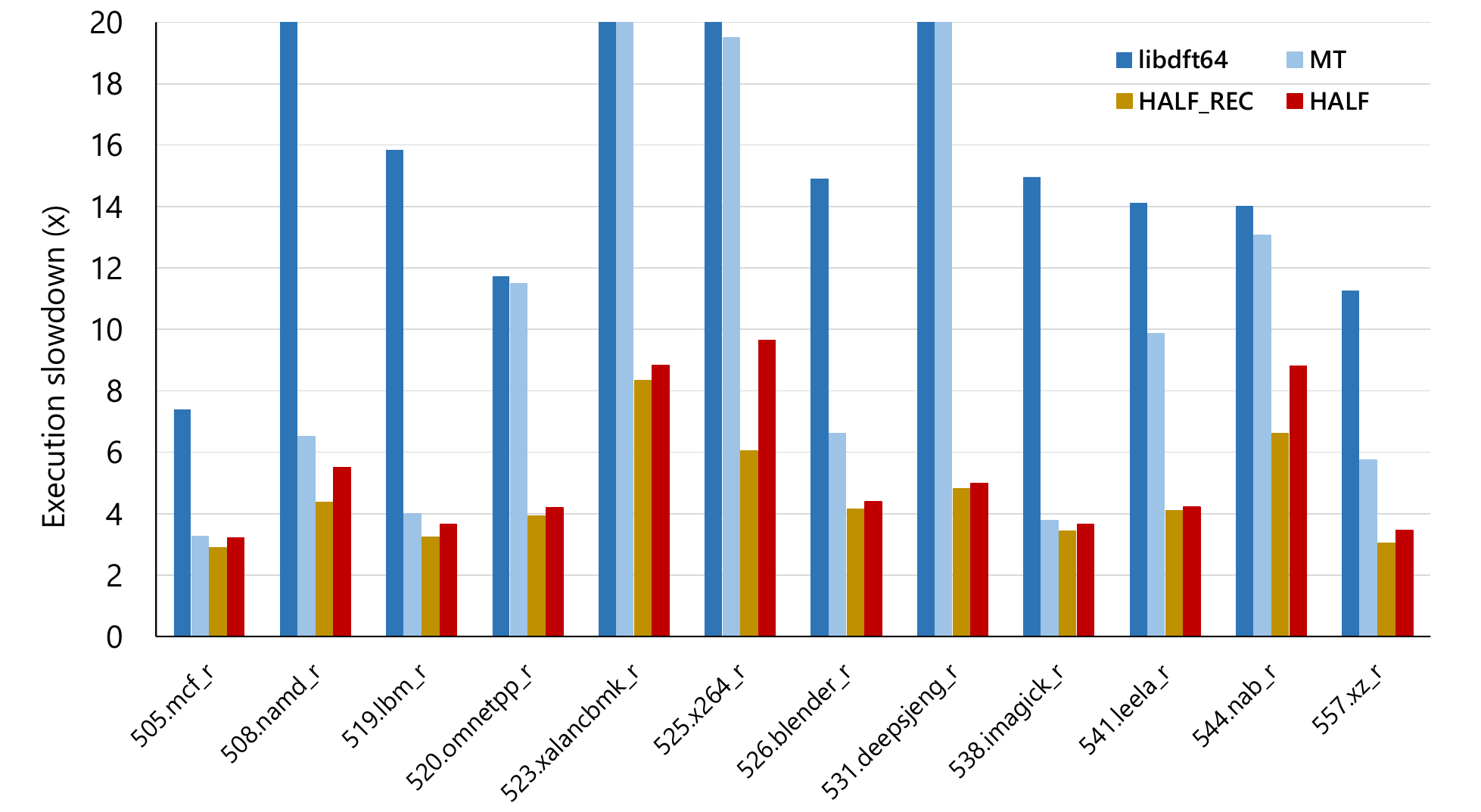}
\caption{End-to-end overhead on SPEC CPU2017.
We compare instruction-grain DTA across \textit{libdft64} (in-process), DynamoRIO memtrace (baseline trace recorder), HALF_REC (recording only), and full HALF analysis.
\textit{libdft64} incurs prohibitive slowdowns (truncated at 20$\times$), while HALF keeps overhead close to simple tracing and adds only marginal cost beyond recording, illustrating the efficiency of exception-driven decoupling.}
\label{fig:perfover2}
\end{figure}

While HALF prioritizes transparency and isolation over raw speed, we conducted a comparative analysis against \textit{libdft64}~\cite{port_2020_port,vuzzer64libdft64_2020_vuzzer64libdft64} (a ported 64-bit version of the classic \textit{libdft}) to demonstrate its practical viability.
\textit{libdft64} represents the state-of-the-art for in-process DTA but suffers from significant overhead due to heavy instrumentation.

We implement taint propagation rules in HALF similar to \textit{libdft}: the target records the required execution stream (e.g., memory addresses and control-flow metadata), while the container replays and performs propagation on a same-address view.
We also measured \textbf{HALF_REC}.
This configuration has the target recording data while the analysis thread effectively “skips” processing, which isolates the cost of the decoupled engine itself.

The results in Fig.~\ref{fig:perfover2} demonstrate a substantial performance advantage.
\textit{libdft64} frequently incurs slowdowns exceeding 20$\times$ (our chart cutoff), spiking to over 50$\times$ for \textit{x264_r}.
In contrast, HALF keeps overheads within a practical range, often comparable to the lightweight \textit{memtrace} tool.
Crucially, the gap between full HALF and HALF_REC stays small.
This shows the producer (target) and consumer (container) overlap well, and our exception-driven commit/switch protocol keeps the recorder from stalling when the analysis thread catches up slowly.
Put another way, the marginal cost of propagation mostly hides behind buffered recording instead of being paid directly on the target’s hot path.
That differs from \textit{libdft64}, where propagation logic lives in-process, inflates code size, and updates metadata on every instruction, so the target pays the full slowdown.
The resulting performance gap, therefore, reflects an architectural ceiling rather than an engineering shortfall: in-process designs must fight for the same CPU resources used by the application, and slowdowns grow with analysis complexity.

The memory-access picture makes this even clearer.
Traditional tools pay a constant shadow tax on every load/store, so each access doubles the pressure on the cache and the Translation Lookaside Buffer (TLB).
HALF moves propagation to container threads that can run on other cores with their own caches, and the shared record buffer mostly behaves like a streaming writer/reader that modern prefetchers handle easily.
As a result, the decoupled design breaks the linear dependency between application accesses and metadata work: the target mainly pays for raw data generation and rare buffer-management events while the container absorbs the rest.
This change is not merely an optimization but an architectural shift necessary for practical, layout-sensitive tracking.
The few slowdowns we still see (e.g., \textit{x264_r} and \textit{nab_r}) trace back to Streaming SIMD Extensions (SSE) and Advanced Vector Extensions (AVX)-heavy code that touches much more shadow state.
As with any asynchronous setup, a large producer/consumer imbalance (for example, many threads burstily producing records) can increase analysis lag.
HALF mitigates that by forcing early commits at synchronization syscalls (Section~\ref{sec:kernel}), but eliminating lag would require reintroducing tighter coordination and hurting throughput.

\begin{table}[!htbp]
\centering
\caption{Memory overhead comparison for shadow usage.}
\label{table:compare}
\begin{tabular}{|>{\centering\arraybackslash}m{3.5cm}|r|r|}
\hline
\multicolumn{1}{|>{\centering\arraybackslash}m{3.5cm}|}{\textbf{Program}} &
\multicolumn{1}{c|}{\textbf{libdft64 (MB)}} &
\multicolumn{1}{c|}{\textbf{HALF (MB)}} \\
\hline
505.mcf_r & 950 & 2.5 \\
508.namd_r & 208 & 4.0 \\
519.lbm_r & 426 & 2.6 \\
520.omnetpp_r & 261 & 6.2 \\
523.xalancbmk_r & 404 & 5.4 \\
525.x264_r & 203 & 4.8 \\
526.blender_r & 233 & 13.4 \\
531.deepsjeng_r & 719 & 3.0 \\
538.imagick_r & 44 & 3.4 \\
541.leela_r & 82 & 3.4 \\
544.nab_r & 56 & 3.0 \\
557.xz_r & 914 & 3.0 \\
\hline
\end{tabular}
\small{
\\
\textit{Note:} Values exclude base DBI tool overhead.
HALF dynamically maps shadow memory, whereas libdft statically reserves large regions.}
\end{table}

Table~\ref{table:compare} highlights a second, transparency-critical advantage: \textbf{shadow-state placement}.
In-process tools like \textit{libdft64} must reserve and manage large shadow-memory regions inside the target address space (often hundreds of MBs), which both increases memory pressure and creates a structural risk of address conflicts.
HALF instead maps the shadow state in the container and materializes it on demand; the target process does not pay for these long-lived reservations and therefore preserves its native allocation trajectory.
This explains the order-of-magnitude footprint gap in Table~\ref{table:compare}: HALF’s container-side shadow state stays at a few megabytes (MBs) for these runs, while \textit{libdft64}‘s in-process reservation grows to hundreds of megabytes (MBs) and can approach a gigabyte (GB).
Here, MB refers to megabytes.
Mechanistically, this is the physical manifestation of our address-space isolation goal: moving shadow memory out of the target eliminates both memory competition and layout deformation rather than attempting to ``negotiate’’ for remaining space.

Table ~\ref{table:ds1} reports fine-grained statistics for SPEC runs under full HALF analysis (default 512KB per-thread record buffers), including instruction counts, analysis-code footprint, and page-fault behavior.
The container’s memory usage (AM) largely follows the target’s actual allocation and access patterns because shadow states are mapped on demand.
Buffer fills reach the million scale (BF), yet the exception-driven switching does not dominate overhead.
Container-side page faults (CF), primarily from on-demand shadow mapping, are typically below the pages committed by the target (TC), indicating that HALF materializes only the shadow pages that are actually accessed.
Collectively, the low AM and manageable fault counts (TF/CF) in Table~\ref{table:ds1} mechanistically validate that same-address container execution eliminates target-side shadow reservations while keeping exception-driven coordination off the hot path.
This confirms that the overhead trends in Fig.~\ref{fig:perfover2} are attributable to removing instruction-level branching and in-process propagation, rather than benchmark-specific artifacts.

\begin{table*}[!htbp]
\centering
\caption{Data statistics for benchmark program analysis experiments.}
\label{table:ds1}
\begin{tabular}{|>{\centering\arraybackslash}m{3.6cm}|r|r|r|r|r|r|r|r|}
\hline
\multicolumn{1}{|>{\centering\arraybackslash}m{3.6cm}|}{\textbf{Program}} &
\multicolumn{1}{c|}{\textbf{Basic Blocks (k)}} &
\multicolumn{1}{c|}{\textbf{Instructions (k)}} &
\multicolumn{1}{c|}{\textbf{PI (\%)}} &
\multicolumn{1}{c|}{\textbf{AM (MB)}} &
\multicolumn{1}{c|}{\textbf{BF (m)}} &
\multicolumn{1}{c|}{\textbf{TF (k)}} &
\multicolumn{1}{c|}{\textbf{TC (k)}} &
\multicolumn{1}{c|}{\textbf{CF (k)}} \\
\hline
505.mcf_r & 11 & 54 & 37 & 1.0 & 1.6 & 79 & 326 & 216 \\
508.namd_r & 20 & 146 & 37 & 2.5 & 1.1 & 72 & 57 & 40 \\
519.lbm_r & 11 & 58 & 36 & 1.1 & 0.6 & 46 & 105 & 105 \\
520.omnetpp_r & 51 & 237 & 39 & 4.7 & 1.7 & 101 & 151 & 59 \\
523.xalancbmk_r & 41 & 201 & 44 & 3.9 & 3.1 & 133 & 199 & 92 \\
525.x264_r & 25 & 178 & 38 & 3.3 & 1.6 & 60 & 56 & 40 \\
526.blender_r & 74 & 406 & 35 & 7.4 & 5.4 & 94 & 91 & 35 \\
531.deepsjeng_r & 15 & 76 & 35 & 1.5 & 2.6 & 56 & 176 & 179 \\
538.imagick_r & 21 & 102 & 35 & 1.9 & 1.3 & 40 & 6 & 5 \\
541.leela_r & 19 & 97 & 39 & 1.9 & 3.3 & 301 & 1039 & 8 \\
544.nab_r & 15 & 79 & 35 & 1.5 & 2.4 & 45 & 18 & 10 \\
557.xz_r & 13 & 76 & 36 & 1.5 & 0.6 & 62 & 284 & 225 \\
\hline
\end{tabular}
\vspace{10pt}
\\ The percentage of instructions requiring instrumentation for recording (PI), analysis code memory usage (AM), number of times record buffers are filled during analysis (BF), page faults during target program execution (TF), pages committed during target program execution (TC), and page faults generated by container processes (CF) are reported.
\end{table*}

\subsubsection{Buffer Switching}

\begin{figure}[!htbp]
\centering
\includegraphics[width=\linewidth]{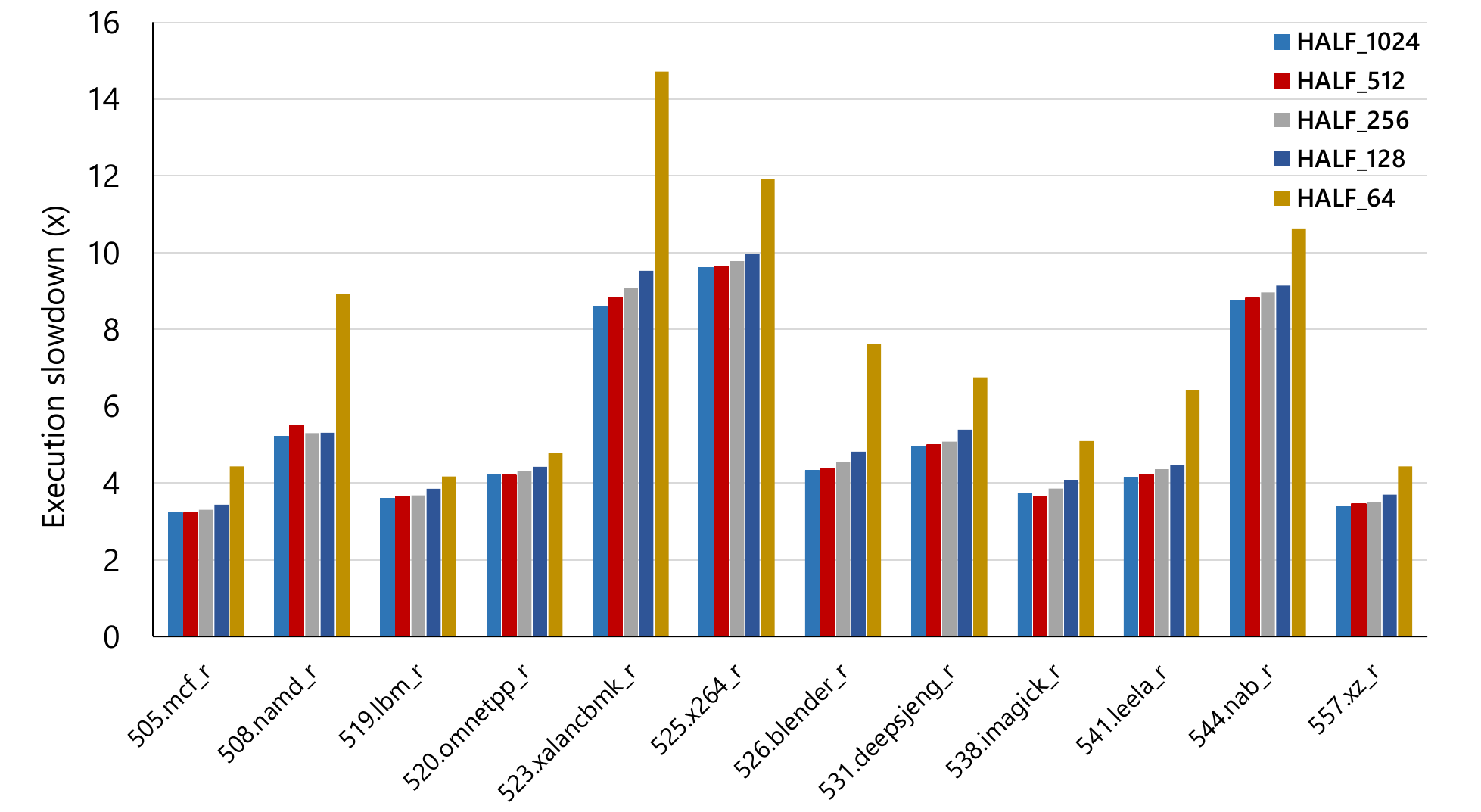}
\caption{Overhead for varying per-thread record-buffer sizes on SPEC CPU2017.
A 64KB buffer increases slowdown due to frequent guard-page faults and buffer rotations.
Overhead stabilizes beyond 128KB; our prototype uses 512KB by default.}
\label{fig:perfover4}
\end{figure}

Record-buffer size trades off exception frequency and memory footprint.
We evaluate several buffer sizes on SPEC CPU2017 (Fig.~\ref{fig:perfover4}).
A small 64KB buffer causes a visible performance cliff for memory-intensive workloads because guard-page faults and buffer rotations become frequent enough that kernel entry/exit overhead starts to dominate the savings from streamlined recording.
This behavior directly supports the amortization model of our exception-driven design: when buffers are too small, the constant cost per exception is paid too often; when buffers are sufficiently large, the same constant cost is spread over many recorded events and becomes negligible.
Once the buffer size reaches 128KB, overhead stabilizes, and further increases provide diminishing returns, indicating that buffer-switch exceptions have returned to the ``rare’’ regime that zero-check recording relies on.
In practice, we recommend choosing the smallest buffer size at or above 128KB to balance memory footprint and exception frequency; our prototype default is 512KB.
It is worth noting that this tuning exposes a practical trade-off under extreme thread counts: per-thread buffers can accumulate into nontrivial memory usage.
HALF mitigates this via early virtual reservation and fixed-size slicing (Section~\ref{sec:memory_communication}), but deployments should still tune buffer counts/sizes to match target concurrency and available memory.

\subsection{Case Studies}
To see how HALF behaves on real software, we ran DTA on six network-facing utilities: \textit{aria2c} (1.36), \textit{certutil} (Windows built-in), \textit{curl} (8.4), \textit{links} (2.25), \textit{wget} (1.21), and \textit{pscp} (0.79).
These programs cover multi-threading, dynamic loading, encrypted I/O, and a variety of module counts.

To keep the workload consistent, we hosted a Hypertext Transfer Protocol Secure (HTTPS) file server in a local VM with Transport Layer Security (TLS) enabled and pulled a 10MB file over each tool, \textit{pscp}, which used Secure Shell (SSH).
We treated the receive path (instrumenting \textit{NtDeviceIoControlFile}) as the taint source and checked taint before file writes via \textit{NtWriteFile}.
Each experiment was repeated 10 times, and the reported numbers are averages.

\subsubsection{Effectiveness and Coverage}

\begin{figure}[!htbp]
\centering
\includegraphics[width=\linewidth]{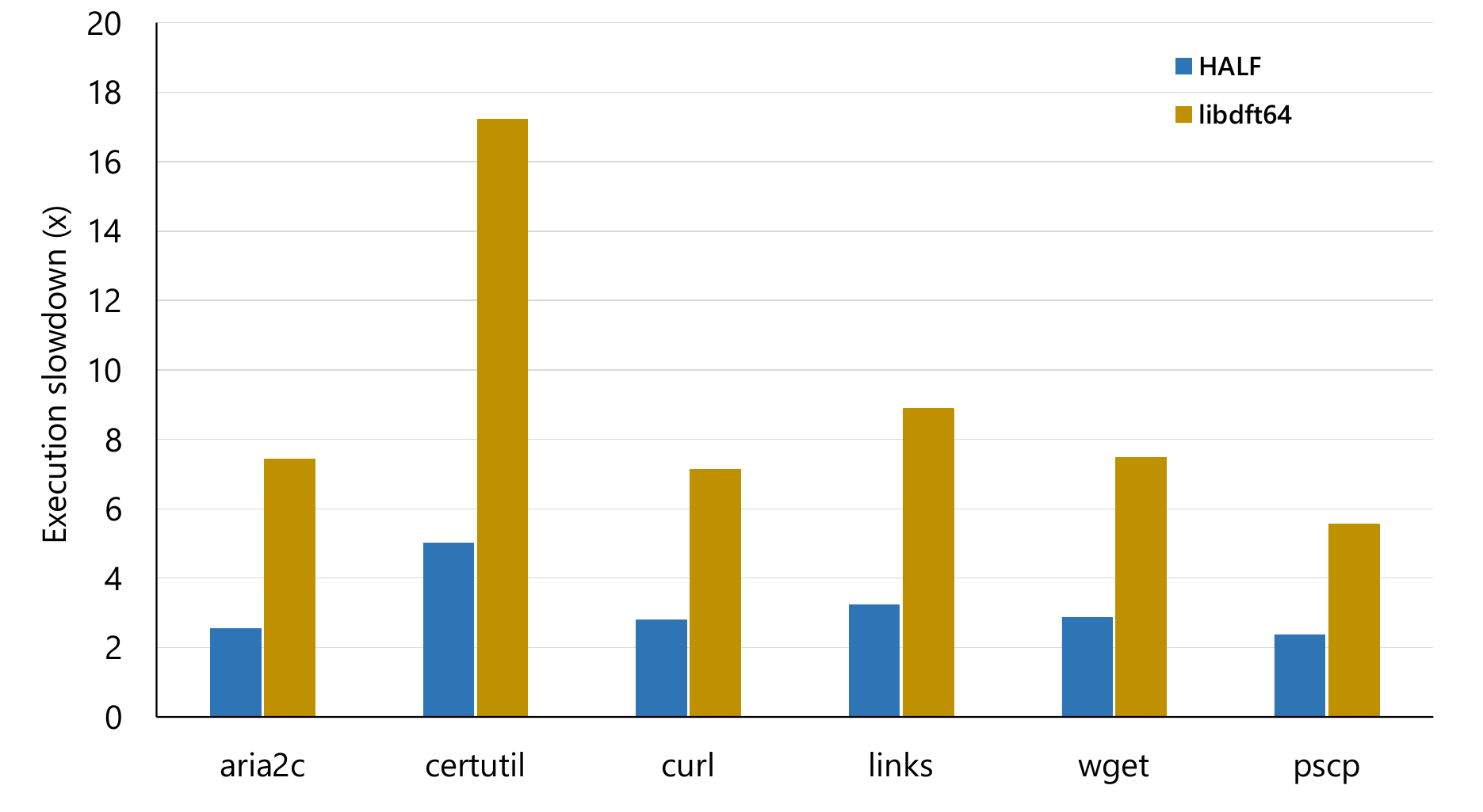}
\caption{Slowdown on six real applications under HALF and \textit{libdft64}.
HALF introduces a few-$\times$ overhead across workloads, while \textit{libdft64} incurs substantially higher slowdowns, demonstrating the practicality of decoupled execution on complex real programs.}
\label{fig:anatime}
\end{figure}

Fig.~\ref{fig:anatime} reports slowdown ratios for these applications.
HALF sticks to a modest 2$\times$–5$\times$ overhead, while \textit{libdft64} slows down by much larger factors, even though both apply the same taint-tracking logic.
That gap traces back to the decoupled architecture: the target simply records stream data and executes lightweight stubs, while the container handles heavyweight propagation off the hot path.
Table~\ref{table:ds2} drills down further.
Even when module counts climb into the tens (and almost eighty for \textit{certutil}), the AM stays in the single-digit MB range, and CF stays low, which reflects on-demand mapping of shadow state rather than eager pre-allocation.
Put differently, HALF’s memory usage tracks what the target actually touches instead of reserving worst-case blobs.
That said, these case studies focus on network receive sources and file-write sinks, so they do not cover other semantics like GUI input, IPC channels, or kernel-object taints.
Extending to those sources simply requires adding the relevant instrumentation on top of the same hollowing and same-address machinery.

\begin{table*}[!htbp]
\centering
\caption{Data statistics for real program analysis experiments}
\label{table:ds2}
\footnotesize
\begin{tabular}{|>{\centering\arraybackslash}c|r|r|r|r|r|r|r|r|}
\hline
\multicolumn{1}{|>{\centering\arraybackslash}c|}{\textbf{Program}} &
\multicolumn{1}{c|}{\textbf{Thread Count}} &
\multicolumn{1}{c|}{\textbf{Module Count}} &
\multicolumn{1}{c|}{\textbf{Basic Blocks (k)}} &
\multicolumn{1}{c|}{\textbf{Instructions (k)}} &
\multicolumn{1}{c|}{\textbf{PI (\%)}} &
\multicolumn{1}{c|}{\textbf{AM (MB)}} &
\multicolumn{1}{c|}{\textbf{TF (k)}} &
\multicolumn{1}{c|}{\textbf{CF (k)}} \\
\hline
aria2c & 2 & 42 & 47 & 224 & 40 & 4.3 & 45 & 1.0 \\
certutil & 8 & 79 & 107 & 502 & 37 & 9.2 & 68 & 14.3 \\
curl & 2 & 28 & 44 & 203 & 37 & 4.1 & 43 & 0.7 \\
links & 8 & 41 & 56 & 265 & 38 & 5.1 & 66 & 4.1 \\
wget & 2 & 30 & 43 & 204 & 36 & 4.1 & 43 & 0.8 \\
pscp & 2 & 37 & 39 & 182 & 38 & 3.6 & 42 & 1.0 \\
\hline
\end{tabular}
\vspace{10pt}
\end{table*}

Table ~\ref{table:ds3} reports task-level statistics for these runs, including end-to-end analysis time (AT) and byte-level taint coverage (RB/CB/DB).

\begin{table*}[!htbp]
\centering
\caption{Data statistics for real program DTA}
\label{table:ds3}
\begin{tabular}{|>{\centering\arraybackslash}m{3.8cm}|r|r|r|r|r|r|r|}
\hline
\multicolumn{1}{|>{\centering\arraybackslash}m{3.8cm}|}{\textbf{Program}} &
\multicolumn{1}{c|}{\textbf{AT (s)}} &
\multicolumn{1}{c|}{\textbf{BF}} &
\multicolumn{1}{c|}{\textbf{RT}} &
\multicolumn{1}{c|}{\textbf{WT}} &
\multicolumn{1}{c|}{\textbf{RB}} &
\multicolumn{1}{c|}{\textbf{CB}} &
\multicolumn{1}{c|}{\textbf{DB}} \\
\hline
\multirow{2}{*}{aria2c} & 2.8 & 3735 & 2110 & 1115 & A09B5E & A00000 & A00000 \\
\cline{2-8}
& 23.2 & 1721 & 850 & 734 & A09B5E & A00000 & A00000 \\
\hline
\multirow{2}{*}{certutil} & 5.0 & 625 & 890 & 2562 & A0A2F3 & 1400000 & 1400000 \\
\cline{2-8}
& 25.3 & 581 & 788 & 2562 & A0A2F3 & 1400000 & 1400000 \\
\hline
\multirow{2}{*}{curl} & 2.6 & 237 & 3471 & 1280 & A09ACB & A00000 & A00000 \\
\cline{2-8}
& 23.0 & 362 & 2998 & 1280 & A09ACB & A00000 & A00000 \\
\hline
\multirow{2}{*}{links} & 3.3 & 601 & 3469 & 112 & A09B5E & A00000 & A00000 \\
\cline{2-8}
& 23.7 & 601 & 2796 & 209 & A09B5E & A00000 & A00000 \\
\hline
\multirow{2}{*}{wget} & 2.7 & 460 & 3764 & 1280 & A09B5E & A00000 & A00000 \\
\cline{2-8}
& 23.1 & 529 & 3265 & 1280 & A09B5E & A00000 & A00000 \\
\hline
\multirow{2}{*}{pscp} & 3.0 & 2272 & 2560 & 2578 & A00000 & A32B8C & A32A8C \\
\cline{2-8}
& 23.4 & 2040 & 2560 & 2574 & A00000 & A151E1 & A1500A \\
\hline
\end{tabular}
\vspace{10pt}
\\ Analysis time (AT), number of data packets received during analysis (RT), number of file writes during analysis (WT), total bytes marked as taint when receiving data packets (RB), total bytes checked before writing to files (CB), and total bytes detected as taint (DB) are reported.
RB/CB/DB are reported in hexadecimal (bytes).
\end{table*}

To stress longer executions, we throttled both upload and download bandwidth to 4 Mbps and reran the experiments.
The second row for each program in Table ~\ref{table:ds3} corresponds to the throttled runs.
While RT and BF vary due to different I/O scheduling, the taint-coverage results remain consistent: for most workloads, RB/CB/DB remain identical across runs.
This is a semantic validation of cross-process tracking correctness: despite different schedules and synchronization interleavings, the same-address container observes the same pointer values and the same taint-propagation outcomes, indicating that decoupling did not break the evidence chain.

Table ~\ref{table:ds3} also reports byte-level coverage.
RB is the total number of bytes marked as tainted at network receive, CB is the number of bytes checked before writing to disk, and DB is the number of bytes that are still tainted at the check.
Because encrypted protocols include handshake and framing, RB can exceed the payload bytes ultimately written to the output file.
Across runs, HALF consistently identifies tainted data reaching file writes; \textit{certutil} shows larger counts because it performs multiple write passes.
For \textit{pscp}, RB and CB/DB are close to the 10MB transferred payload (with additional protocol overhead), indicating that the file-write buffers largely originate from network input and that taint remains coherent through the SSH stack in a multi-module setting.

\begin{table*}[!htbp]
\centering
\caption{Analysis synchronization evaluation for real programs}
\label{table:dssync}
\begin{tabular}{|>{\centering\arraybackslash}m{3.5cm}|r|r|r|r|r|}
\hline
\multicolumn{1}{|>{\centering\arraybackslash}m{3.5cm}|}{\textbf{Program}} &
\multicolumn{1}{c|}{\textbf{WSN}} &
\multicolumn{1}{c|}{\textbf{SSN}} &
\multicolumn{1}{c|}{\textbf{CFN}} &
\multicolumn{1}{c|}{\textbf{DFN}} &
\multicolumn{1}{c|}{\textbf{GSR}} \\
\hline
\multirow{2}{*}{aria2c} & 2319 & 39 & 4 & 0 & 100\% \\
\cline{2-6}
& 261 & 41 & 5 & 0 & 100\% \\
\hline
\multirow{2}{*}{certutil} & 3743 & 1744 & 1076 & 5 & 99.5\% \\
\cline{2-6}
& 1229 & 943 & 297 & 4 & 98.7\% \\
\hline
\multirow{2}{*}{curl} & 2478 & 26 & 2 & 0 & 100\% \\
\cline{2-6}
& 1112 & 27 & 2 & 0 & 100\% \\
\hline
\multirow{2}{*}{links} & 54867 & 34416 & 4007 & 732 & 81.7\% \\
\cline{2-6}
& 16980 & 11537 & 1219 & 156 & 87.2\% \\
\hline
\multirow{2}{*}{wget} & 32 & 22 & 1 & 0 & 100\% \\
\cline{2-6}
& 31 & 24 & 1 & 0 & 100\% \\
\hline
\multirow{2}{*}{pscp} & 2689 & 37 & 4 & 0 & 100\% \\
\cline{2-6}
& 901 & 39 & 5 & 0 & 100\% \\
\hline
\end{tabular}
\vspace{10pt}
\\
Wait-related system call invocations (WSN), signal-related system call invocations (SSN), total number of buffers requiring checking at checkpoints for all threads (CFN), total number of marked buffers detected by all threads (DFN), and global synchronization rate (GSR) are reported.
\end{table*}

\subsubsection{Synchronization Evaluation}
We measure how well HALF’s event-driven commits keep decoupled analysis in sync for multi-threaded programs.
Reconstructing precise synchronization relationships without heavy static analysis is hard, so we rely on the Global Synchronization Rate (GSR) as a simpler proxy.
The idea is to mark buffers committed by a signaling thread at the signal point (Fig.~\ref{fig:atsync}, position A) and clear the mark once the corresponding analysis thread processes that buffer.
When another thread hits a related checkpoint and commits (position B), we check whether any outstanding signal marks remain.
CFN counts these checkpoint checks, and DFN counts how often they see outstanding marks.
We define $GSR=\frac{CFN-DFN}{CFN}$, so higher values mean signal-related buffers usually finish before the next checkpoint.

Table~\ref{table:dssync} shows that most programs reach near-100\% GSR, which means actual analysis progress stays timely for typical workloads.
This directly validates the kernel-mediated syscall interception mechanism: forcing early commits at wait/signal syscalls injects low-frequency, semantics-relevant synchronization points that cap analysis lag without constant polling or heavy IPC.
High GSR also suggests that kernel-enforced barriers keep the happens-before relationships needed for correct taint propagation at synchronization edges.
Even the most concurrency-heavy case (\textit{links}) with tens of thousands of wait/signal syscalls and thousands of checkpoint checks maintains 81.7\% GSR (and 87.2\% under throttling), which shows syscall-filtered commits tame lag in realistic multi-threaded I/O.
Without kernel-level forced commits, the producer could keep producing records while the consumer lags, and DFN would grow because dependent threads would reach checkpoints before signal-related buffers are processed.
That counterfactual highlights why the observed high GSR is evidence that synchronization is enforced at the right place (kernel syscalls) instead of being approximated in user space.
This reliance on syscall interception has a boundary, though: workloads that mostly use pure user-space synchronization primitives (e.g., spin-locks or lock-free queues) expose fewer kernel-visible commits, so the synchronization granularity degrades from syscall-bounded’’ to buffer-capacity-bounded’’ in those cases.

\subsection{Exploit Analysis: Baseline Failure vs HALF Success}
\label{sec:exploit}
We now show that HALF succeeds on layout-sensitive exploits where traditional user-space tools commonly fail.
Heap corruption exploits (heap overflow, Use-After-Free (UAF), etc.) typically build precise layouts through heap spraying, so any analyzer that disturbs that layout will break the payload.
Table~\ref{table:typicalvul} lists the public exploits we tested, covering browsers and document readers.
We compare HALF against \textit{libdft} (an in-process taint tracker that keeps shadow state inside the target) and a fixed-shadow setup inspired by \textit{AddressSanitizer (ASan)}.
The baseline failures stem from \textbf{address-space conflict}; this is not an implementation bug but an unavoidable consequence of keeping the shadow state in the same process as the target.
Take CVE-2023-21608 (Acrobat Reader): its PoC sprays allocations around a fixed low-address range (\texttt{0x40000000}–\texttt{0x80000000}) so crafted objects land at predictable offsets before the use-after-free.
An in-process tracker such as libdft reserves a large contiguous shadow region at a fixed address band, and when those two constraints overlap, the target allocator cannot get the intended spray layout.
Allocations fail, land elsewhere, or shift subsequent objects, and the exploit diverges or crashes before the payload ever runs.
This example illustrates why \textbf{kernel-assisted process hollowing} is necessary: by rebuilding the container’s address space to mirror the target, HALF preserves the native layout and avoids the structural conflicts that plague in-process approaches.
The hollowed container, therefore, keeps the exploit’s heap spraying on its native \textbf{allocation trajectory}, which is essential for triggering the vulnerability.

To be specific, the Adobe Reader exploit relies on the Windows Low Fragmentation Heap (LFH) to service a sequence of allocations that predictably land around the \texttt{0x40000000} region.
In the \textit{libdft} baseline, the static reservation of shadow memory fragments the virtual address space early in the process lifecycle.
This can force the LFH to select alternative heap segments for the spray, disrupting the relative offsets between sprayed objects and the vulnerable buffer.
Consequently, when the UAF is triggered, the control flow is redirected to a location that contains uninitialized data rather than the intended return-oriented programming chain, leading to an immediate access violation instead of code execution.

Thus, hollowing is not merely an implementation choice, but a prerequisite for ensuring the scientific validity of dynamic analysis on layout-sensitive adversarial binaries.

The exploits in Table ~\ref{table:typicalvul} are publicly available PoCs for memory-corruption vulnerabilities in browsers and document readers; all targets are analyzed as binaries.

\begin{table*}[!htbp]
\centering
\caption{Analysis results for typical exploit programs.}
\label{table:typicalvul}
\footnotesize
\begin{tabular}{|>{\centering\arraybackslash}m{4.0cm}|c|c|c|c|c|c|}
\hline
\multirow{2}{*}{\textbf{CVE ID}} & \multirow{2}{*}{\textbf{Vulnerability Type}} & \multirow{2}{*}{\textbf{Target Program}} & \multirow{2}{*}{\textbf{Version}} & \multicolumn{3}{c|}{\textbf{Payload Can Execute Successfully}} \\
\cline{5-7}
& & & & \textbf{libdft} & \textbf{ASan} & \textbf{HALF} \\
\hline
CVE-2017-11882 & Memory Corruption & Microsoft Office 16.0.4266.1001 & 32-bit & \Checkmark & \Checkmark & \Checkmark \\
CVE-2018-11529 & Use-After-Free & VLC Media Player 2.2.8 & 64-bit & \XSolidBrush & \XSolidBrush & \Checkmark \\
CVE-2020-26950 & Use-After-Free & Firefox 71.0 & 64-bit & \XSolidBrush & \XSolidBrush & \Checkmark \\
CVE-2021-30632 & Out-of-Bounds Write & Chrome 91.0.4472.124 & 64-bit & \Checkmark & \XSolidBrush & \Checkmark \\
CVE-2022-28672 & Use-After-Free & Foxit PDF Reader 11.1.0 & 32-bit & \XSolidBrush & \Checkmark & \Checkmark \\
CVE-2023-21608 & Use-After-Free & Acrobat Reader 2022.003.20258 & 32-bit & \XSolidBrush & \XSolidBrush & \Checkmark \\
\hline
\end{tabular}
\end{table*}

\begin{table*}[!htbp]
\centering
\caption{Data statistics for exploit analysis experiments}
\label{table:dsvul}
\footnotesize
\setlength{\tabcolsep}{2pt} % 保持较小间距

% 关键修改：把 c 改为 Y，这样表头太长时会自动换行，而不是撑爆表格
\begin{tabularx}{\linewidth}{|*{9}{Y|}}
\hline
\textbf{Program} &
\textbf{Thread Count} &
\textbf{Module Count} &
\textbf{Basic Blocks (k)} &
\textbf{Instructions (k)} &
\textbf{Time (s)} &
\textbf{PI (\%)} &
\textbf{BF (k)} &
\textbf{TC (MB)} \\
\hline
VLC & 11 & 94 & 2 & 6 & 12 & 44 & 2.8 & 1610 \\
Office & 52 & 138 & 763 & 3530 & 29 & 65 & 6.7 & 450 \\
Acrobat & 14 & 160 & 550 & 2550 & 20 & 68 & 24 & 1720 \\
Chrome & 7 & 40 & 30 & 160 & 3 & 50 & 0.02 & 16 \\
Firefox & 19 & 72 & 25 & 135 & 16 & 49 & 121 & 2500 \\
\hline
\end{tabularx}
\end{table*}

Table ~\ref{table:typicalvul} summarizes whether each exploit payload executes successfully under three setups: (i) in-process \textit{libdft}, (ii) a fixed-shadow layout similar to \textit{ASan}, and (iii) HALF.
We treat successful payload execution (opening the Calculator) as a proxy that the exploit’s required layout was preserved under analysis.
Because our targets are closed-source binaries, the ASan baseline uses a fixed shadow-memory reservation scheme rather than recompilation.
For multi-process browsers, we attach instrumentation only to the renderer process to keep experiments controlled.
HALF preserves payload execution in all cases, while \textit{libdft} and the fixed-shadow baseline fail on most layout-sensitive exploits due to address-space conflicts introduced by large shadow reservations.
Failures are more common for 32-bit targets where the virtual address space is tighter.

Not all exploits are strongly layout-sensitive.
CVE-2017-11882 succeeds across all configurations.
In contrast, the ASan-like scheme succeeds on CVE-2022-28672 only because its default reservation (\texttt{0x20000000}) does not overlap the PoC’s spray range (approximately \texttt{0x19b40000}).
In real samples, spray ranges are unpredictable, fixed-shadow schemes are brittle, and in 64-bit settings, they can also collide with runtime allocations (e.g., TEB/stack) performed by the OS and instrumentation runtime.

Table ~\ref{table:dsvul} reports runtime statistics for the exploit runs.
Some exploits allocate large amounts of memory (e.g., Acrobat/Firefox), which increases the likelihood of conflicts for in-process baselines.
For browsers, we again focus on renderer processes; the Chrome PoC happens to succeed under \textit{libdft} due to smaller allocation pressure, while the Firefox PoC does not.
We emphasize that payload executability is used as a proxy for layout preservation; it does not by itself prove full semantic soundness of taint propagation under all exploit-induced races.
In addition, multi-process exploit chains may require instrumenting multiple cooperating processes; HALF supports this, but our controlled setup focuses on the renderer to keep the counterfactual comparison (address conflicts) isolated.

\subsection{Malware Analysis}
We further validate HALF using a realistic malware workflow.
This experiment involves the analysis of the Cobalt Strike \textit{Artifact} payload (\texttt{artifact.exe}).
In a typical stageless run, the loader decrypts and executes shellcode, downloads a Beacon module, reflectively loads the module into memory, and eventually jumps into the Beacon code through indirect branches.
Common examples of such branches include instructions like \textit{call eax}.
This workflow places stress on two specific areas where decoupled systems typically struggle.
The first challenge is the creation of dynamically generated or executable pages at runtime.
The second challenge involves pointer-rich control transfers where targets must align across domains.
HALF addresses both issues through its same-address semantics.
Specifically, the hollowed container shares the virtual layout of the target.
Consequently, a pointer observed by the target remains valid within the container and allows for inspection without the need for address translation or rebasing.
Since HALF operates at the virtual-memory and page-fault levels using standard OS mechanisms, we executed the same prototype on Windows 10 (versions 20H2 and 21H1) and observed stable behavior.

\begin{figure}[!t]
\centering
\includegraphics[width=\linewidth]{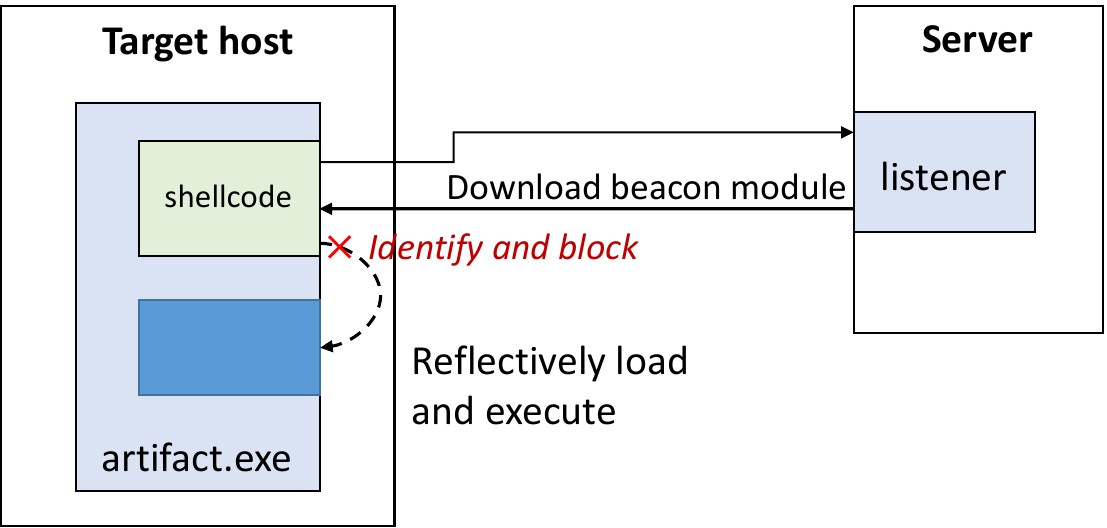}
\caption{\normalsize The workflow for Beacon module loading and detection under HALF.
HALF taints buffers received from the network and validates indirect \textit{call} or \textit{jmp} targets.
When a control transfer targets tainted memory, such as the downloaded Beacon code, HALF reports the event.
The system can optionally terminate execution to maintain a controlled analysis environment.}
\label{fig:beaconload}
\end{figure}

We taint buffers received over the network by instrumenting the Ancillary Function Driver receive path, which includes functions such as \textit{NtDeviceIoControlFile}.
Additionally, we insert a lightweight task that validates the targets of indirect \textit{call} or \textit{jmp} instructions.
If a target points into tainted memory, HALF reports a potential in-memory execution attempt (Fig.~\ref{fig:beaconload}).
The stub on the target side records only the branch target.
Meanwhile, the container performs the taint lookup and reports on the same-address view.
This design keeps the hot path minimal and adheres to our principle of ``record fast, analyze outside.‘’

For this experiment, we built the client using Cobalt Strike 4.4 and hosted the server in a local VM.
HALF identifies the tainted control transfer within approximately 150\ ms after the arrival of the first packet.
This detection occurs before the Beacon module finishes loading.
The sample receives approximately 35 packets and taints roughly 256 KB of memory.
Analysis shows that about 90\% of the first 512 bytes checked at the indirect transfer are tainted, which demonstrates that the downloaded code is tracked end-to-end.
The malware creates approximately ten threads and loads roughly 38 modules.
This behavior is similar to that of a normal application.
If we disable the optional post-detection termination and allow the run to continue, the Beacon successfully connects to the server.
This result indicates that HALF maintains low runtime overhead in this scenario.
However, this design entails a trade-off regarding transparency.
Kernels that contain loaded drivers or exception handling mechanisms can leave detectable footprints.
Sophisticated adversaries might attempt to enumerate drivers or use timing-based probes to detect the system (Section~\ref{sec:limitations}).
Nonetheless, HALF improves transparency compared to in-process trackers because it removes large shadow reservations, even though the kernel monitor introduces a visible attack surface.

\section{Discussion}

\subsection{The Necessity of Structural Isolation}
Conventional wisdom in dynamic analysis relies on the assumption that instrumentation remains transparent if it is sufficiently ``thin.‘’
Our findings challenge this assumption within the context of adversarial binaries.
We posit that Address Space Invariance serves as a prerequisite for the analysis of layout-sensitive threats.

Any tool that performs in-process analysis acts as an observer.
This observer inevitably perturbs the state of the target process.
The most notable perturbation occurs in the memory layout.
For robust software, this perturbation is negligible.
However, exploits that rely on precise heap geometry or malware that employs anti-hooking checks are sensitive to these changes.
Small layout shifts can cause execution to collapse into a failure mode, such as failed spraying or divergent control flow.
The use of process hollowing in HALF represents more than an engineering maneuver to conserve memory; it constitutes a methodological shift to restore the isolation between the observer and the target.
By moving the analysis state, which functions as the observer,‘’ into a hollowed container and enforcing same-address semantics, we ensure that the target’’ perceives a near-native environment.
This approach resolves the fundamental conflict between high-fidelity observation and layout preservation.

\subsection{Portability and Compatibility}
Although our prototype operates on Windows, the core mechanism is portable.
This mechanism relies on kernel-assisted process hollowing.
The requirements for implementation are threefold: 1) the ability to intercept process creation, 2) the ability to manipulate virtual memory mappings, such as unmapping and remapping, and 3) the ability to catch page faults.
Linux, for instance, provides \texttt{ptrace} and \texttt{userfaultfd}.
Developers could combine these features to achieve similar ``zero-check’’ recording, although kernel-level remapping might still require a loadable kernel module to ensure performance.
We prioritized Windows due to the scarcity of effective fine-grained analysis tools for its closed-source ecosystem.

Our kernel monitor relies on documented kernel callbacks, such as \texttt{PsSetCreateProcessNotifyRoutine}, and standard memory management APIs.
This reliance minimizes instability.
We validated the system on Windows 10 versions 20H2 and 21H1.
Since the driver does not hook internal kernel structures, such as regions protected by PatchGuard, it maintains high stability.
However, the reliance on a kernel driver implies that deployment requires administrative privileges.
Furthermore, modern systems impose signing requirements, such as test-signing or valid certificates.

\subsection{Security Risks}
HALF is designed for analysis rather than online protection.
The introduction of a kernel driver inherently expands the attack surface.
An attacker who complies with the interface of the driver could potentially manipulate container processes.
Therefore, users should deploy HALF in controlled analysis environments, such as sandboxes or analyst workstations.
We mitigate risks by strictly validating IOCTLs and ensuring that the driver operates only on processes designated for analysis.

\subsection{Limitations and Trade-offs}
\label{sec:limitations}
We acknowledge several boundaries in the current design and propose potential directions for future mitigation.

\noindent\textbf{Synchronization Granularity.}
Asynchronous analysis inevitably introduces lag between execution and analysis.
We mitigate this issue via exception-driven commits at syscalls related to synchronization (Fig.~\ref{fig:atsync}).
However, race conditions in highly concurrent targets or pure user-space synchronization primitives, such as spin locks and lock-free queues, may still lead to transient under-taint or over-taint because they do not trigger kernel intervention.
To address this, future iterations could incorporate \textit{heuristic spin-loop detection}, where the instrumentation module identifies repeated execution of short basic blocks and forces a buffer commit.
Alternatively, static analysis could pre-identify custom synchronization routines to insert explicit barrier stubs, thereby restoring happens-before relationships without intercepting every instruction.

\noindent\textbf{Anti-Analysis and Invisibility.}
HALF eliminates user-mode artifacts, such as unexpected injected DLLs or conspicuous address-space holes, which are easy for malware to detect.
Despite this, it introduces indicators at the kernel level.
A sophisticated adversary could enumerate loaded drivers or measure page-fault latency (timing side channels) to infer the presence of the monitor.
However, such detection typically requires visibility at the kernel level or high-precision timing.
These requirements raise the bar for evasion compared to the difficulty of detecting user-mode hooks.
To further harden the system against kernel-level probing, the kernel monitor could be migrated to a lightweight hypervisor (Ring -1), leveraging hardware virtualization extensions (e.g., Intel Virtualization Technology (VT-x)).
This architectural shift would render the monitor invisible to standard kernel queries and allow the hypervisor to mask timing anomalies introduced by page faults.

\noindent\textbf{Resource Bounds.}
For extremely memory-intensive applications, such as those with working sets larger than 10GB, the dual-process model increases memory pressure.
This model consists of the Target and the Container.
On-demand mapping reduces unnecessary physical commitments, but it cannot bypass limits on system-wide physical memory.
To alleviate this pressure, we propose implementing \textit{shadow memory compression}.
Since shadow data often exhibits high redundancy (e.g., large contiguous regions of ``untainted’’ status), compressing these pages within the container could significantly reduce the physical footprint.
Furthermore, an adaptive mechanism could degrade analysis granularity from byte-level to object-level when system memory reaches critical thresholds, ensuring continuity of analysis at the cost of precision.

\noindent\textbf{Instrumentation Artifacts.}
While HALF successfully offloads analysis logic and data, the lightweight instrumentation engine, DBI, itself must reside within the target process.
Although we minimize its footprint, advanced malware might still detect the presence of the DBI code cache or the just-in-time compilation delay.
Future work seeks to decouple the system further by replacing the in-process DBI recorder with hardware-assisted tracing features, such as PT.
This approach would allow the target to run natively without any injected code, while the container consumes the hardware trace stream, achieving the ultimate goal of zero-artifact analysis.

\section{Conclusion}
This paper presented HALF as a fine-grained dynamic analysis framework for Windows that resolves the conflict between transparency and efficiency.
HALF preserves the native address-space layout of target programs through kernel-assisted process hollowing.
The system keeps target-side instrumentation lightweight while it executes tracking logic in a synchronized, same-address container.
A lightweight kernel monitor orchestrates this architecture to minimize the impact on the target execution.
Experimental results involving SPEC CPU2017, real-world applications, and exploits demonstrate that HALF reduces runtime overhead relative to in-process baselines.
It also circumvents the memory allocation conflicts that typically cripple traditional tools.
As binary threats become more environmentally sensitive, the structural isolation provided by HALF offers a robust foundation for future defensive research.
We plan to enhance synchronization soundness for concurrent workloads and explore the portability of these mechanisms to other operating systems like Linux.

\section*{Acknowledgments}
This work is supported by
the Major Program of the National Natural Science Foundation of China under Grant 62293503,
the Major Research Plan of the National Natural Science Foundation of China under Grant 92467202,
the Frontier Technologies R\&D Program of Jiangsu under Grant BF2024071,
the Open Fund of Anhui Province Key Laboratory of Cyberspace Security Situation Awareness and Evaluation under Grant TK224013
and the Postgraduate Research and Practice Innovation Program of Jiangsu Province under Grant KYCX24\_1231.

\bibliographystyle{IEEEtran}
\bibliography{ref}

\begin{IEEEbiography}[{\includegraphics[width=1in,height=1.25in,clip,keepaspectratio]{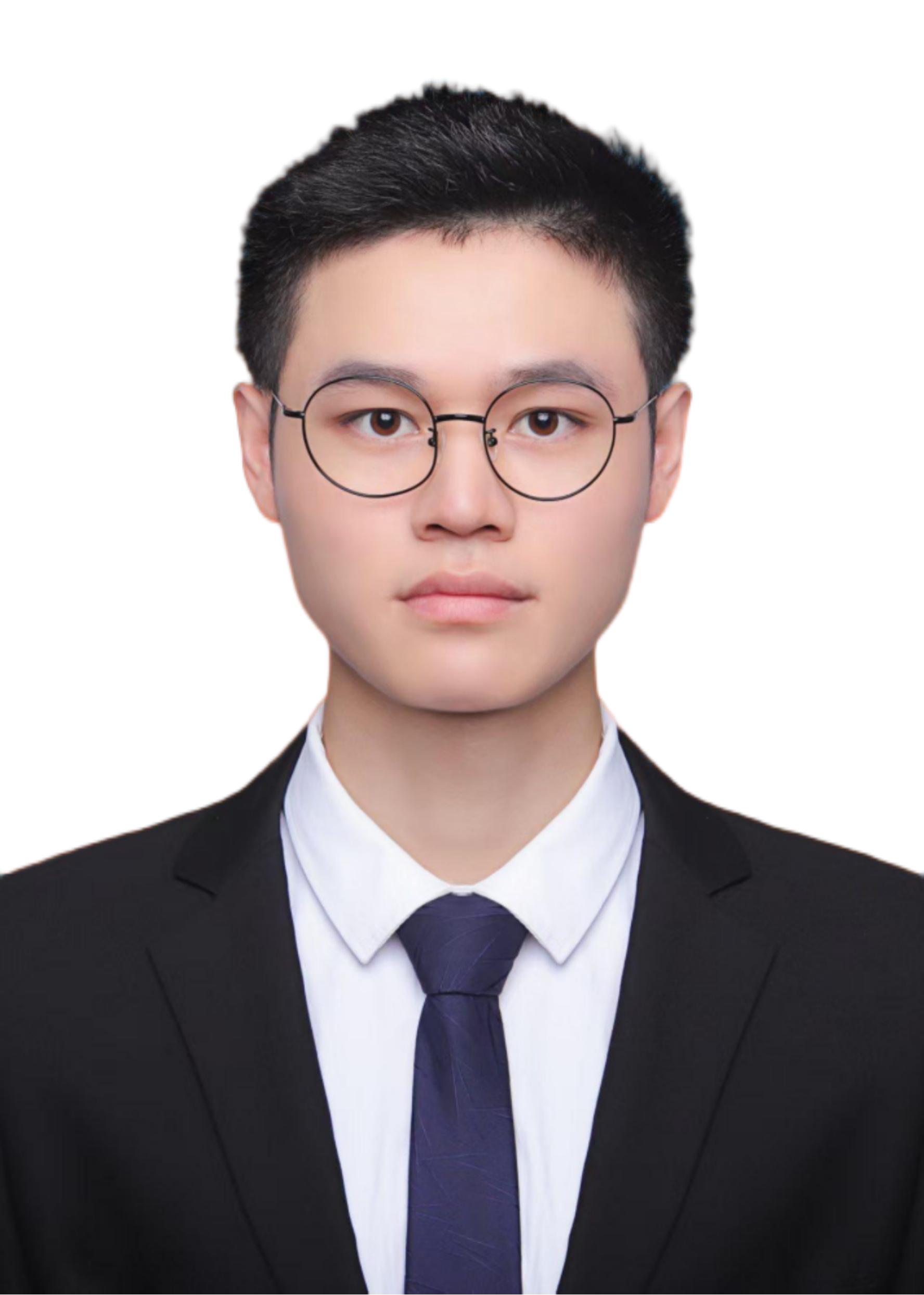}}]{Zhangbo Long}
is currently pursuing his Ph.D.
at the School of Cyberspace Security, Nanjing University of Posts and Telecommunications, Nanjing, China.
His research focuses on system security and malware analysis.\end{IEEEbiography}

\begin{IEEEbiography}[{\includegraphics[width=1in,height=1.25in,clip,keepaspectratio]{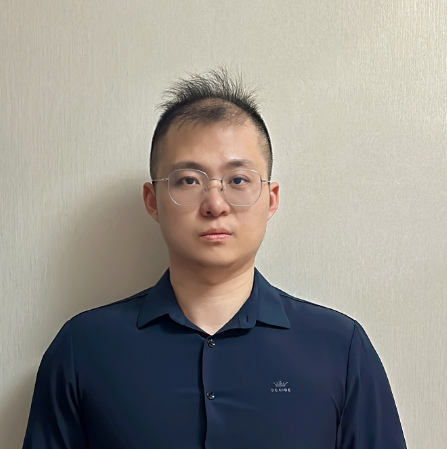}}]{Letian Sha}
received Ph.D.
degree from Wuhan University in 2014.
Currently Professor at the School of Computer Science, Nanjing University of Posts and Telecommunications.
Research interests include vulnerability discovery, malware analysis, and software security.\end{IEEEbiography}

\begin{IEEEbiography}[{\includegraphics[width=1in,height=1.25in,clip,keepaspectratio]{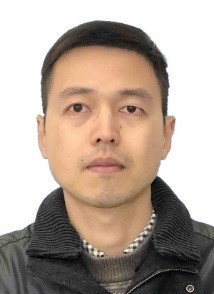}}]{Jiaye Pan}
received Ph.D.
degree from Nanjing University of Aeronautics and Astronautics in 2020.
Currently an Assistant Professor at the School of Computer Science, Nanjing University of Posts and Telecommunications.
Research interests include system security, software security, and network security.\end{IEEEbiography}

\begin{IEEEbiography}[{\includegraphics[width=1in,height=1.25in,clip,keepaspectratio]{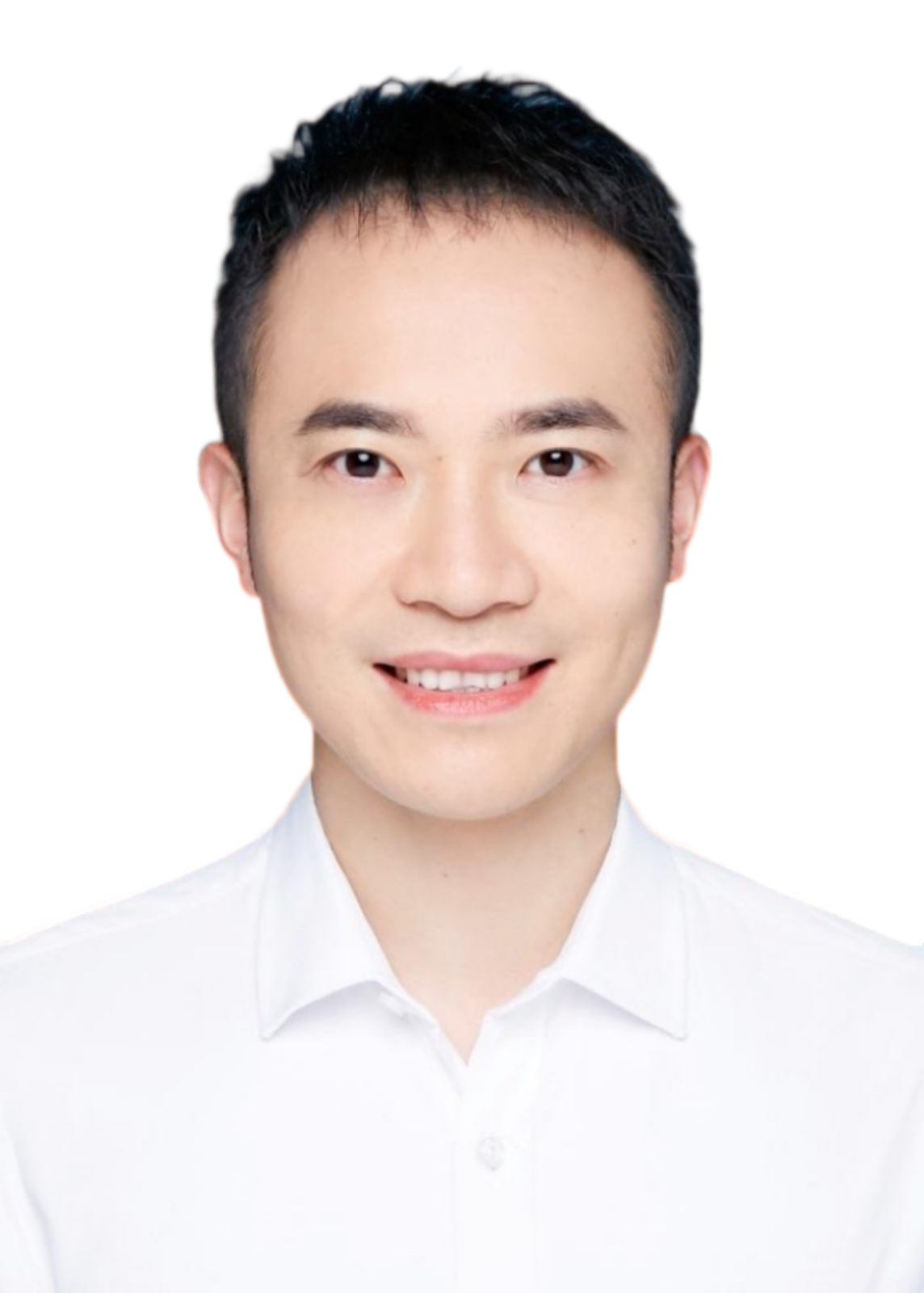}}]{Haiping Huang}
(Member, IEEE) received the B.E.
degree in management science and engineering and the M.E.
degree in computer science and technology from Nanjing University of Posts and Telecommunications, Nanjing, China, in 2002 and 2005, respectively, and the Ph.D.
degree in computer application technology from Soochow University, Suzhou, China, in 2009.
He is currently a Professor and a Ph.D.
Supervisor with the School of Computer Science, Nanjing University of Posts and Telecommunications.
His research interests include information security.\end{IEEEbiography}

\begin{IEEEbiography}[{\includegraphics[width=1in,height=1.25in,clip,keepaspectratio]{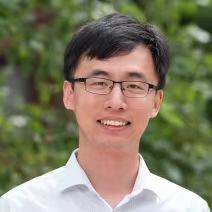}}]{Dongpeng Xu} received Ph.D.
degree from Pennsylvania State University in 2018.
Currently Assistant Professor at the Department of Computer Science, The University of New Hampshire.
Research interests in software security, particularly binary program analysis, software obfuscation/deobfuscation, malware analysis, program similarity analysis, and model checking.
\end{IEEEbiography}

\begin{IEEEbiography}[{\includegraphics[width=1in,height=1.25in,clip,keepaspectratio]{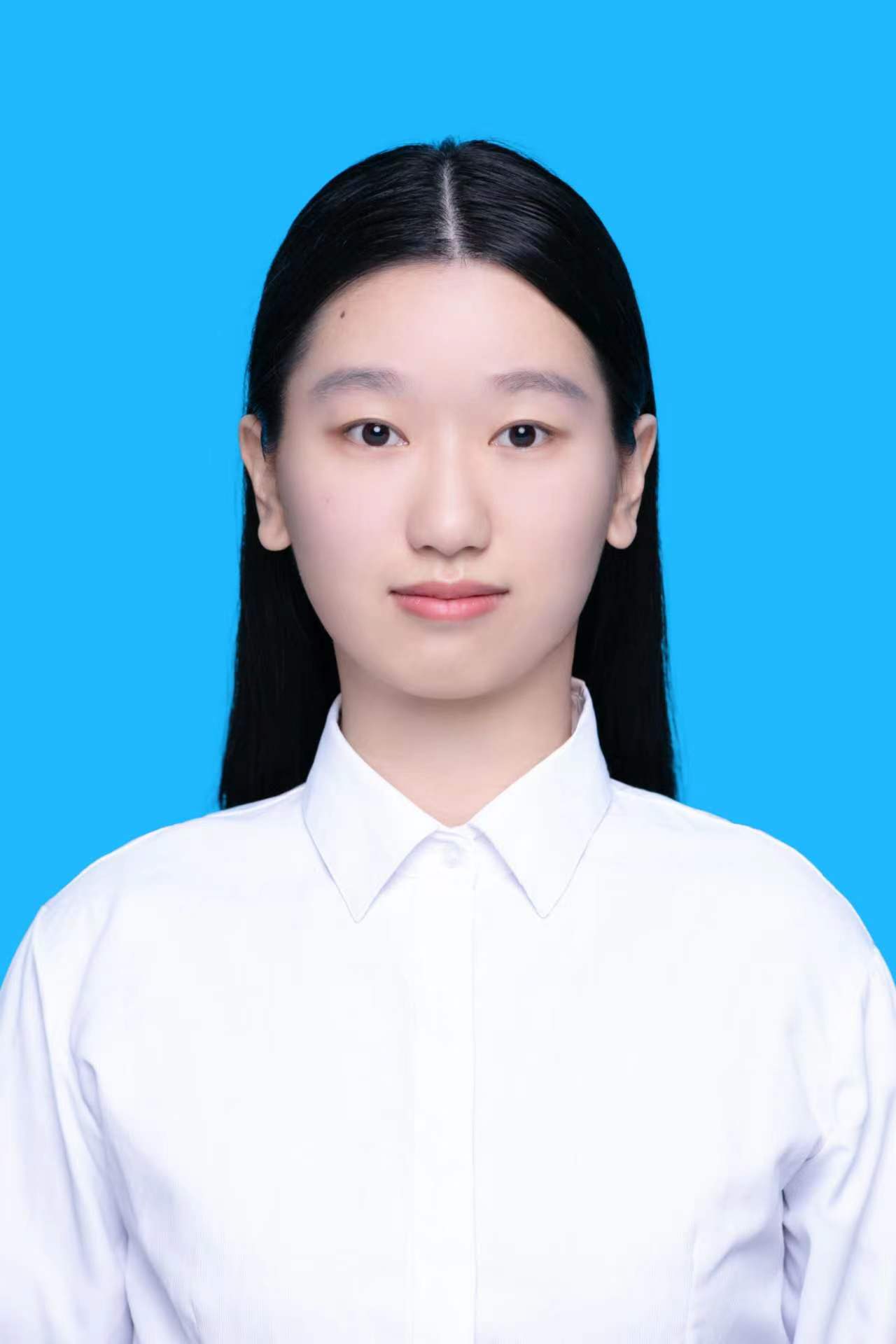}}]{Yifei Huang}
received a Bachelor’s degree in Management from Beijing University of Posts and Telecommunications and an Engineering degree from Queen Mary University of London.
Currently pursuing a Master’s degree in system security at the School of Computer Science, Nanjing University of Posts and Telecommunications.
Research interests include binary security, fuzzing, and software analysis.
\end{IEEEbiography}

\begin{IEEEbiography}[{\includegraphics[width=1in,height=1.25in,clip,keepaspectratio]{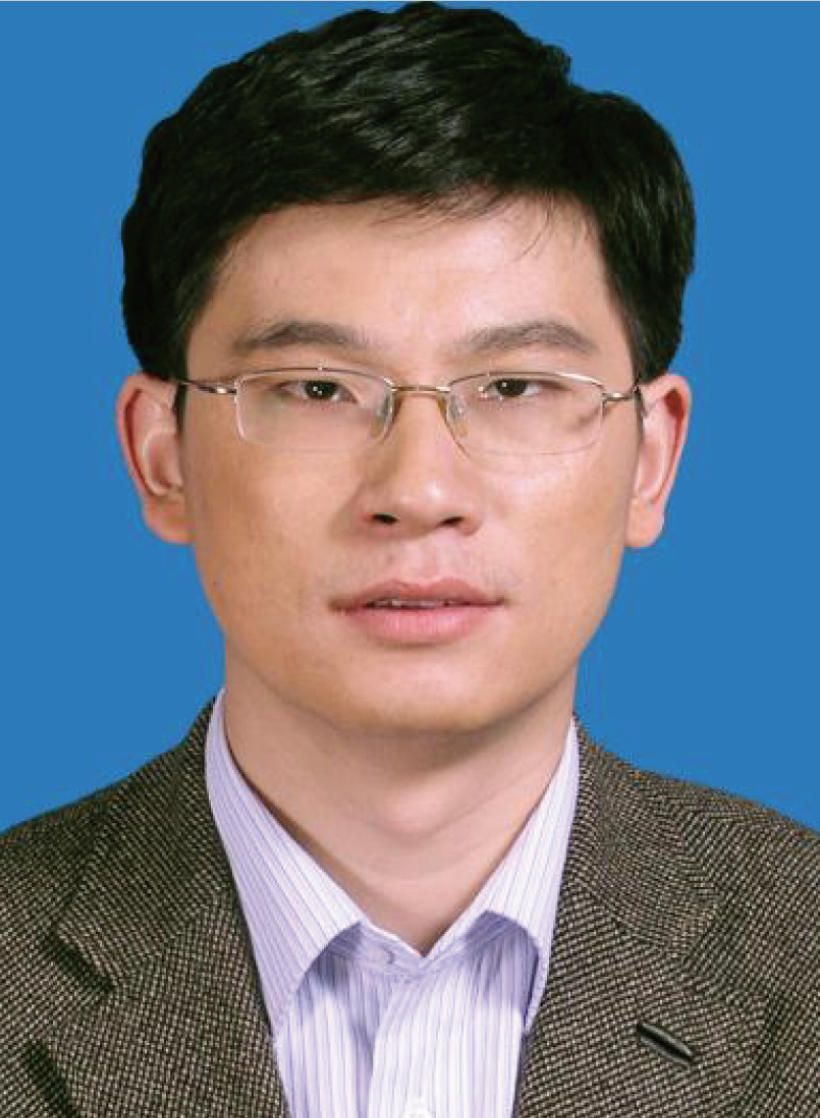}}]{Fu Xiao}
received Ph.D.
degree from Nanjing University of Science and Technology in 2007.
Currently Professor and Ph.D.
Supervisor at the School of Computer Science, Nanjing University of Posts and Telecommunications.
Research results published in important conferences and journals, including INFOCOM, IEEE/ACM TON, IEEE JSAC, IEEE TC, IEEE TPDS, IEEE TDSC, IEEE TMC, ACM TECS, IEEE TVT, etc.
Research interests include wireless sensor networks, Internet of Things security, and information security.
Professor Fu Xiao is an IEEE Computer Society and Association for Computing Machinery (ACM) member.
\end{IEEEbiography}

\vfill

\end{document}